# The implications of outcome truncation in reproductive medicine RCTs: a simulation platform for trialists and simulation study.

Jack Wilkinson [1*], Jonathan Huang [2], Antonia Marsden [1], Michael Harhay[3], Andy Vail [1], Stephen A Roberts [1].

*corresponding author: jack.wilkinson@manchester.ac.uk

[1] Centre for Biostatistics, Manchester Academic Health Science Centre, Division of Population Health, Health Services Research and Primary Care, University of Manchester, UK, M13 9PL.

[2] Singapore Institute for Clinical Sciences, Agency for Science, Technology, and Research, National University of Singapore, Singapore.

[3] Department of Biostatistics, Epidemiology and Informatics, Perelman School of Medicine, University of Pennsylvania, USA.

# Abstract

**Background:** Randomised controlled trials in reproductive medicine are often subject to outcome truncation, where the study outcomes are only defined in a subset of the randomised cohort. Examples include birthweight (measurable only in the subgroup of participants who give birth) and miscarriage (which can only occur in participants who become pregnant). These outcomes are typically analysed by making a comparison between treatment arms within the subgroup (for example, comparing birthweights in the subgroup who gave birth, or miscarriages in the subgroup who became pregnant). However, this approach does not represent a randomised comparison when treatment influences the probability of being observed (i.e. survival). The practical implications of this for the design and interpretation of reproductive trials is unclear however.

**Methods:** We developed a simulation platform to investigate the implications of outcome truncation for reproductive medicine trials. We used this to perform a simulation study, in which we considered the bias, Type 1 error, coverage, and precision of standard statistical analyses for truncated continuous and binary outcomes. Simulation settings were informed by published assisted reproduction trials.

**Results:** Increasing treatment effect on the intermediate variable, strength of confounding between the intermediate and outcome variables, and presence of an interaction between treatment and confounder were found to adversely affect performance. However, within parameter ranges we would consider to be more realistic, the adverse effects were generally not drastic. For binary outcomes, the study highlighted that outcome truncation may commonly cause separation in smaller studies, where none or all of the participants in a study arm experience the outcome event. This was found to have severe consequences for inferences.

**Conclusion:** We have provided a simulation platform which may be used by researchers in the design and interpretation of reproductive medicine trials subject to outcome truncation, and have used this to conduct a simulation study. The study highlights several key factors which trialists in the field should consider carefully to protect against erroneous inferences. Standard analyses of truncated binary outcomes in small studies may be highly biased, and this has implications for meta-analysis, where this scenario commonly arises.

**Background**

Outcome data are usually unavailable for some participants in an RCT. Most frequently, this is due to loss to follow up or patient withdrawal from the study. However, in many reproductive medicine trials, the availability of a participant's outcome data depends on their status in relation to an intermediate response variable. For example, trials of assisted reproductive technologies (ART) are generally conducted in individuals trying to become pregnant and have babies. In these trials, pregnancy outcomes such as miscarriage (occurring only in the subset of women who become pregnant) and infant outcomes such as birthweight (measurable only in participants who have births) are often of interest. These outcomes cannot be collected in all participants, even if there is no loss to follow up, as they are not observable for everyone in the cohort. This phenomenon has been described as 'truncation (or censoring) due to death' (1), because it often arises in studies where mortality precludes measurement of the outcome variable (2). However, since this form of missing data also occurs in populations where mortality is not a material concern, we use the more general term 'outcome truncation'.

In reproductive medicine trials, outcome data subject to truncation are frequently analysed by making a comparison between study arms in the subset of participants who were not truncated. This is typically done by calculating standard measures of treatment effect (such as an unadjusted mean difference or odds ratio) and performing standard statistical tests (such as a t-test or chi-squared test). These approaches would be valid if the treatment had no effect on the intermediate (censoring) variable. Otherwise, various authors have pointed out that standard analyses of truncated outcome data are subject to a form of selection bias, whereby selecting on intermediate outcomes breaks randomization and therefore biases treatment effect estimates (2-7). Figure 1 shows the conditions under which selection bias due to outcome truncation will, in principle, arise. Outcome truncation also reduces the sample size compared to the full randomised cohort, which is anticipated to impact the precision of the effect estimate (4, 8). Such loss in precision would need to be accounted for during the study design stage to ensure adequately powered studies are pursued.

Some authors have nonetheless argued that a comparison of outcomes in the observable study participants *is* the correct analysis, since this captures the effect in the only group of relevance – those who are at risk (9). What this argument misses is that, to the extent the observed difference is caused by selection-induced confounding rather than by a causal effect of treatment, the transportability of the estimate will be restricted to populations with the same distribution of confounders. As a result, there is a concern that the standard approaches to analysis of truncated outcomes might be misleading. Crucially, important findings in reproductive medicine hinge on analyses of this sort. For example, a recent RCT found that the choice of embryo culture medium used in IVF affected the birthweight of babies born from the treatment (10). However, if culture media affect conception or miscarriage rates differently, then the mean observed birthweights correspond to two different populations which may differ with respect to exposures such as smoking. This might be problematic when applying these trial findings to other populations for which this selection does not exist or differs – e.g. if an adjuvant therapy improves conception rates for all subjects, observed differences between media might no longer be applicable. Further examples can be found in systematic reviews published in Cochrane Gynaecology and Fertility, which sometimes report miscarriage rates for trials in which individuals were randomised prior to conception, using the number of women who became pregnant as the denominator (11-13).

Outcome truncation due to pregnancy loss has been studied in the context of harmful exposures in pregnancy and long-term outcomes of children, using simulation (3, 8) and heuristic argument (14). The impact on the study of birth defects has also recently been considered, using analysis of observational data (15). Although these examples are informative, they are tailored to the investigation of epidemiological questions, and the scenarios they describe are likely to be less relevant for trialists working in ART, since various key parameters (magnitude of intervention effects, event rates, strength of confounding) materially differ in the latter compared to the former context. Moreover, while the relevance of treatment-confounder interactions to outcome truncation has been described (14), their importance has not been empirically evaluated in existing simulation studies. As a result, it is not currently clear whether outcome truncation substantively affects the findings of ART RCTs or their clinical interpretations. A greater understanding of the consequences of outcome truncation would assist in the design of ART RCTs, as well as in the reinterpretation of published trials where outcomes were compared in the uncensored subgroup. Additionally, a characterisation of outcome truncation would be useful for researchers developing analytic methods in this area.

To address this need, we developed a simulation platform in R which can be used by reproductive medicine trialists to aid study design and interpretation in the presence of outcome truncation. We used this platform to investigate the impact of outcome truncation on typical statistical analyses used in assisted reproduction RCTs. We investigated both continuous (e.g. birthweight) and binary (e.g., miscarriage) outcomes subject to truncation, using plausible ranges of parameter values informed by published ART studies.

**Methods**

Simulation study

We developed a simulation platform in R to investigate outcome truncation in two-arm trials where treatment is administered on a single occasion (as opposed to an ongoing regimen), an intervening selection event occurs (*e.g.* conception or live birth), and the study outcome is measured at a single point in time. This reflects the situation found in many reproductive medicine trials. We then used this to conduct a simulation study. The primary aim of this study was to characterise outcome truncation in relation to bias, coverage and Type 1 error of standard analyses, in realistic scenarios corresponding to ART RCTs. Code to reproduce the study, or to conduct novel investigations of outcome truncation, is available at https://osf.io/gzqbr/ .

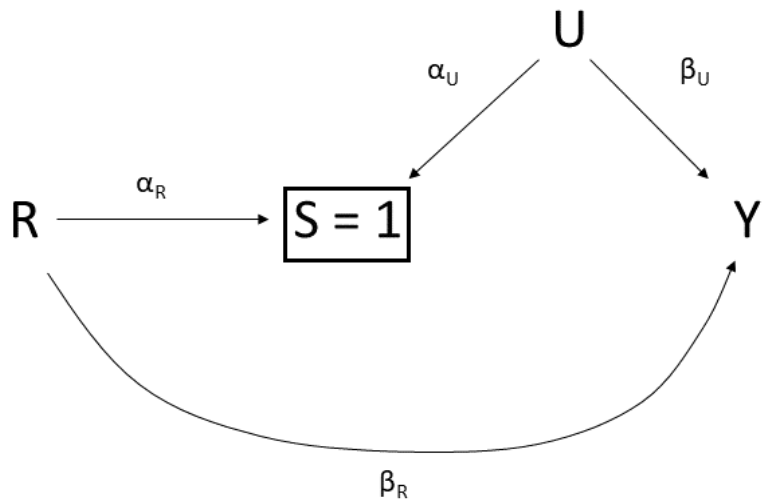

Figure 1: Causal diagram showing conditions that will result in survival bias in an RCT. R = treatment, S = binary intermediate response (e.g. pregnancy, live birth), U = uncontrolled or unmeasured baseline variables, Y = study outcome. Arrows depict causal relationships. For parsimony of presentation, we label the paths with the corresponding parameters from the data generating model used in a simulation study. A box drawn around S=1 indicates that we are conditioning on the intermediate variable: the analysis set includes only participants with S=1. Bias will occur whenever $\alpha_R$, $\alpha_U$, and $\beta_U$ are all nonzero.

We evaluated both a binary and a continuous outcome, and, in the core study, considered two sets of simulations for each. In Set 1, we considered a simple, additive data generating process, without interaction terms (as depicted in Figure 1). In Set 2, we considered the impact of including an interaction between the treatment effect on the intermediate and an unmeasured confounder in the data generating process. Parameter values were informed by several ART RCTs (10, 16) and a recent review of power, precision, and sample size in reproductive medicine studies (17). Table 1 summarises the simulation parameters.

Set 1

For both continuous and binary outcomes, let $i = 1,…, n$ index the $i$th participant with treatment allocation $R_i$ = 1 (treatment) or 0 (control) and intermediate response variable $S_i$ = 0 or 1. We simulated trials with $n$ participants, with $n$ taking values of 100, 200, 500, 1000, divided equally between two study arms. Each patient's probability of having a positive intermediate response was simulated using a logistic model, $\log(\pi_i / (1 − \pi_i )) = \log(0.2) + \alpha_R R_i + \alpha_U u_i$, and the response was then drawn from a Bernoulli($\pi_i$) distribution. The intercept value was selected to correspond to a control group event rate of 17%, which represents a typical live birth rate in an IVF RCT with an unselected population. The treatment effect on the intermediate response variable took values ranging from no effect ($\exp(\alpha_R) = 1$) to very large ($\exp(\alpha_R) = 2$), with the odds ratio (OR) increasing in increments of 0.05. Based on recent work which looked at estimated effect sizes in reproductive medicine meta-analyses (17), we would consider odds ratios larger than 1.2 to be reasonably exceptional. However, we included values larger than this in the simulation study in order to characterise the phenomenon further up to a value of $\exp(\alpha_R) = 5$, corresponding to an implausibly high (in a trial context) effect of treatment on conception or survival probability. This extreme setting was included to provide intuition regarding a plausible upper bound to problems caused by outcome truncation. In particular, we reasoned that if the impact was negligible even in such an extreme scenario, then this would provide some reassurance in relation to real studies.

The patient-specific variable $u_i$ represents prognostic baseline characteristics influencing both the intermediate response variable and the study outcome variable $Y_i$, and was drawn from a Normal (0,1) distribution. We set $\exp(\alpha_U)$ equal to 0.8, such that higher values of the prognostic index resulted in reduced probability of the intermediate response, and therefore reduced probability of having the outcome observed.

In this context, the magnitude of the effect on the intermediate response variable determines the size and degree of size imbalance of the uncensored cohort, and this was anticipated to affect the performance of standard analyses. We purposefully did not correct for this, since this is part of the phenomenon under study.

*Continuous outcomes*

For the study of continuous outcomes, we simulated $Y_i$ from a Normal($\mu_i$, $580^2$) distribution, with $\mu_i$ = 3300 + $\beta_R R_i$ + $\beta_U u_i$. The values for the standard deviation (SD) and intercept were based on the point estimates for the SD and mean for birthweight (in grams) in a recent trial of embryo culture media (10). We set $\beta_U$ to -116, corresponding to -0.2 standard deviations in the outcome. This represents lower outcome values (e.g. reduced birthweight) for participants with higher values of the confounder $u_i$. The coefficient $\beta_R$ corresponds to the effect of treatment allocation on the outcome, excluding any selection effects arising due to a treatment effect on the intermediate response. We considered values for $\beta_R$ ranging from 0 to 2 SDs in increments of 0.1 SDs, with an additional setting of 5 SDs representing an extreme test case. The outcome measurements for the uncensored cohort were then selected by excluding participants who had $S_i$ = 0.

*Binary outcomes*

For the study of binary outcomes, we simulated $p_i$ = Pr($Y_i$ = 1) using a logistic model, log($p_i$ / (1 – $p_i$)) = log(0.1) + $\beta_R R_i$ + $\beta_u u_i$, and then drew $Y_i$ from a Bernoulli($p_i$) distribution. We set exp($\beta_u$) = 1.2, so that a higher value of the confounder $u_i$ corresponded to an increased chance of having the outcome. Recall that we set increasing values of the prognostic index to result in a *lower* probability of the intermediate response occurring – this scenario was chosen to reflect the case where the intermediate response is pregnancy, and the outcome is an adverse pregnancy outcome, such as miscarriage. There may be patient characteristics which make pregnancy less likely, while also reducing the chance that the pregnancy will be carried to term (meaning that miscarriage occurs). The intercept corresponds to an event rate of 9%. The treatment effect $\beta_R$ took values ranging from no effect ($\exp(\beta_R)$ = 1) to very large ($\exp(\beta_R)$ = 2), increasing in increments of 0.05 on the OR scale. A value of 5 was included as a test case. Once again, the outcome measurements for the uncensored cohort were then selected by excluding participants who had $S_i$ = 0.

Set 2

Set 2 was as for Set 1, but with an interaction term $\alpha_{RU}$ between treatment and $u_i$ in the data generating model for the intermediate variable. For both the continuous and binary outcome studies, we simulated $S_i$ from a Bernoulli($\pi_i$) distribution with $\log(\pi_i / (1 - \pi_i)) = \log(0.2) + \alpha_R R_i + \alpha_U u_i + \alpha_{RU} R_i u_i$, with $\exp(\alpha_{RU})$= 0.8.

The simulations were computationally cheap, allowing us to simulate and analyse a relatively large number of datasets corresponding to each tested scenario. The number of iterations per scenario was set to 10,000. Simulations were conducted in R (18), and ggplot2 (19), ggpubr (20) and ggthemes (21) were used for visualisation. Random seeds were obtained from random.org.

| Parameter description | Notation | Values |
|---|---|---|
| Total number of randomised participants | $n$ | 100, 200, 500, 1000 |
| **Set 1** | | |
| Generation of intermediate variable | | |
| Intercept (log(odds) of event in control group) | | $\log_e(0.2)$ |
| Treatment effect on intermediate variable | $\alpha_R$ | $\log_e(1, 1.05, 1.1,\ldots 2, 5)$ |
| Effect of unmeasured confounding on intermediate variable | $\alpha_u$ | $\log_e(0.8)$ |
| Generation of outcome variable – continuous outcome study | | |
| Intercept (control group mean) | | 3300 |
| Treatment effect on outcome | $\beta_R$ | 0, 0.1, 0.2,…,2,5 SDs |
| Effect of unmeasured confounding on outcome variable | $\beta_u$ | -0.2 SDs |
| Generation of outcome variable – binary outcome study | | |
| Intercept (log(odds) of event in control group) | | $\log_e(0.1)$ |
| Treatment effect on outcome | $\beta_R$ | $\log_e(1, 1.05, 1.1,\ldots 2, 5)$ |
| Effect of unmeasured confounding on outcome variable | $\beta_u$ | $\log_e(1.2)$ |
| **Set 2 (differences from Set 1)** | | |
| Generation of intermediate variable | | |
| Interaction between treatment and unmeasured confounding | $\alpha_{RU}$ | $\log_e(0.8)$ |

**Table 1: Summary of parameter values used in core simulation study**

*Sensitivity analyses*

We conducted a number of sensitivity analyses, each of which involved making a uniform change to both Sets 1 and 2. These explored the impact of a) increasing the strength of confounding between the intermediate and outcome variables, b) changing the direction of the treatment effect on the intermediate variable and c) increasing event rates for the intermediate and binary outcome. In a) we increased the strength of the effect of the confounder $u_i$ on the intermediate variable to $\alpha_u = \log(0.5)$. We increased the effect of the confounder on the continuous outcome to $\beta_U = -1SD$ and the effect of the confounder on the binary outcome to be $\beta_U = \log(1.5)$. In b) we considered $\alpha_R = 1/\log_e(1, 1.05, 1.1,…2, 5)$. This was done to check that different influences were not operating in opposing directions, cancelling each other out, and obfuscating performance issues. In c) we increased the intercepts in the intermediate and binary outcome submodels to be $\log(1)$, corresponding to a substantially elevated event rate of 50%.

*Estimand*

In this context, several estimands could be considered. We compared estimates to $\beta_R$, representing the effect of treatment on the outcome variable, in the hypothetical case where no censoring would occur (a hypothetical estimand, in the terminology of recent guidance on estimands in clinical trials (18). We selected this because this corresponds to a common interpretation given to analyses in this context. For example, in a recent trial investigating embryo culture media, a relative decrease in birthweight associated with one medium was interpreted as demonstrating a physiological effect on the embryo and foetus (10), rather than support for the hypothesis that any increase in live birth rate might be associated with worse perinatal outcomes due to selection effects. We explore this point in more detail in the discussion.

*Analysis methods*

In the continuous outcome study, we evaluated the difference in means and associated standard inferential methods (two-sample equal variance t-test and 95% confidence intervals based on the t distribution). In the binary outcome study, we evaluated the sample odds ratio and 95% confidence interval based on the profile likelihood following a logistic regression fit (19). We also evaluated three statistical tests; a chi-squared test, an adjusted 'N-1' chi-squared test (20), and Fisher's exact test. The adjusted chi-squared test involves multiplying the test statistic by (N-1)/N, with N the overall sample size (in this case, the total sample size in the subgroup with outcome data available for analysis), and has been suggested to perform well in small samples (21).

*Performance measures*

In the continuous outcome study, we evaluated bias, coverage, Type 1 error, model SE, and empirical SE (22). In the binary outcome study, we evaluated bias of the log(OR), model SE, empirical SE, and coverage. We also calculated the Type 1 error of the chi-squared test, adjusted chi-squared test, and Fisher's exact test. For binary outcomes, we removed separated instances (those where no participants in a treatment arm experienced the outcome event) before calculating the performance measure, including these in the total counts of instances of missing data.

**Results**

Continuous outcome study

Missing data due to inability to compute estimates from simulated datasets did not prove to be a material problem in the continuous outcome study as the greatest amount of missing data in any scenario was 0.05%.

*Bias*

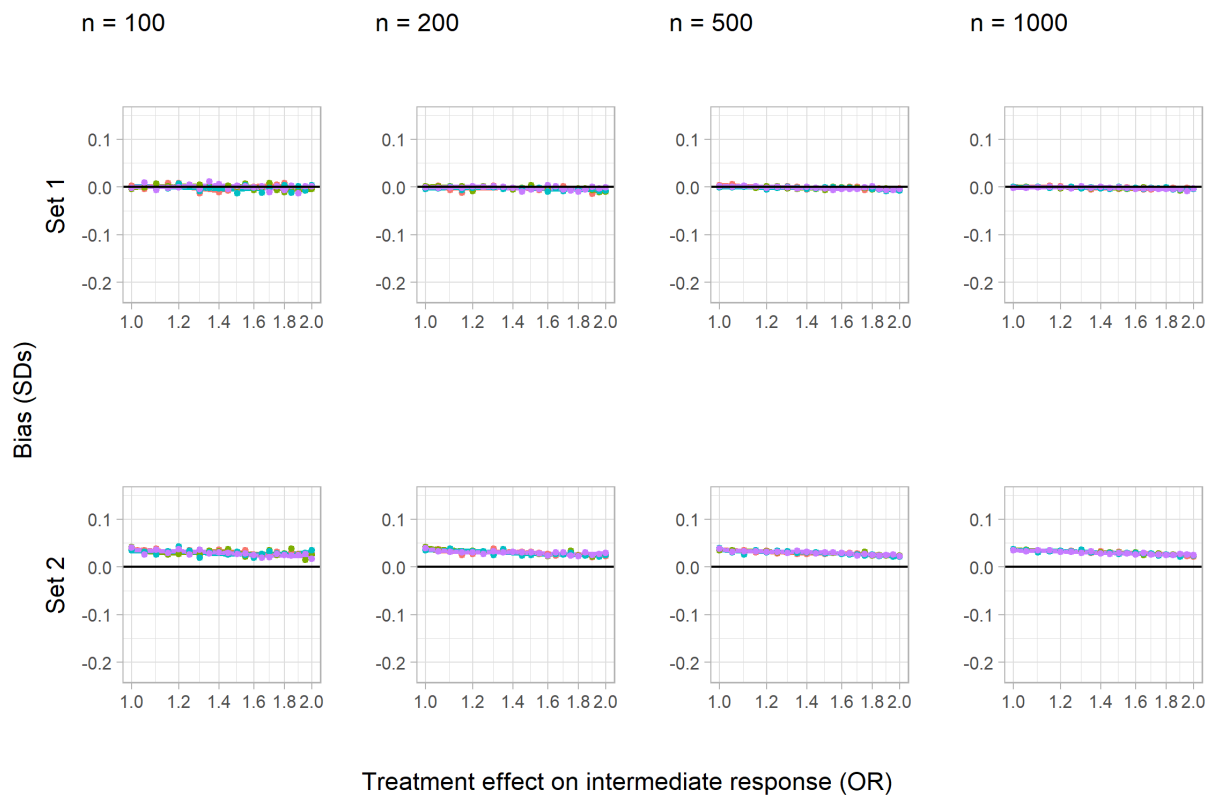

Treatment effect on intermediate response (OR)

**Figure 2: Bias of a simple difference in means in the continuous outcome simulation study (core scenarios). Colour indicates treatment effect on the outcome variable (SDs): red = 0, green = 0.2, blue = 1, purple = 5.**

In the core scenarios, bias was present but very small when there was no interaction between treatment and the intermediate variable (Set 1, Figure 2). In Set 1, bias increased as the treatment effect on the intermediate variable increased, but remained negligible even when this effect became implausibly large; Figure 2 shows this for an OR as large as 2, but in an extreme test case (OR = 5) the bias still did not exceed -0.02SDs. In Set 2 by contrast, which included an interaction between treatment and the unmeasured confounder in the generation of the intermediate variable, bias can be seen to decrease with increasing treatment effect on the intermediate up to an OR of 2, and was very close to zero for an OR of 5. Neither sample size (columns in Figure 2), nor the magnitude of the effect of treatment on the outcome (colours in Figure 2) influenced bias.

Sensitivity analysis A however demonstrated that increasing confounding between the intermediate and the outcome variable modified the impact of the treatment effect on the intermediate variable (slopes steeper in S Figure 1 compared to Figure 2), such that even in the absence of interactions (Set 1) noticeable bias could arise for larger values of the OR. The bias was still reasonably modest for these larger OR values in Set 1, however (below 0.1 SDs), and small for more realistic values of the parameter (below 0.05 SDs for OR < 1.2). In the presence of an interaction (Set 2) bias was substantial for these realistic OR values however.

Changing the direction of the treatment effect on the intermediate in sensitivity analysis B showed that negative ORs did not result in qualitative changes to the relationship with bias (S Figure 2). Increasing the incidence of the intermediate variable in sensitivity analysis c) reduced bias for the set with an interaction, (Set 2), since this meant that a reduced proportion of the cohort was subject to outcome truncation (S Figure 3).

*Coverage*

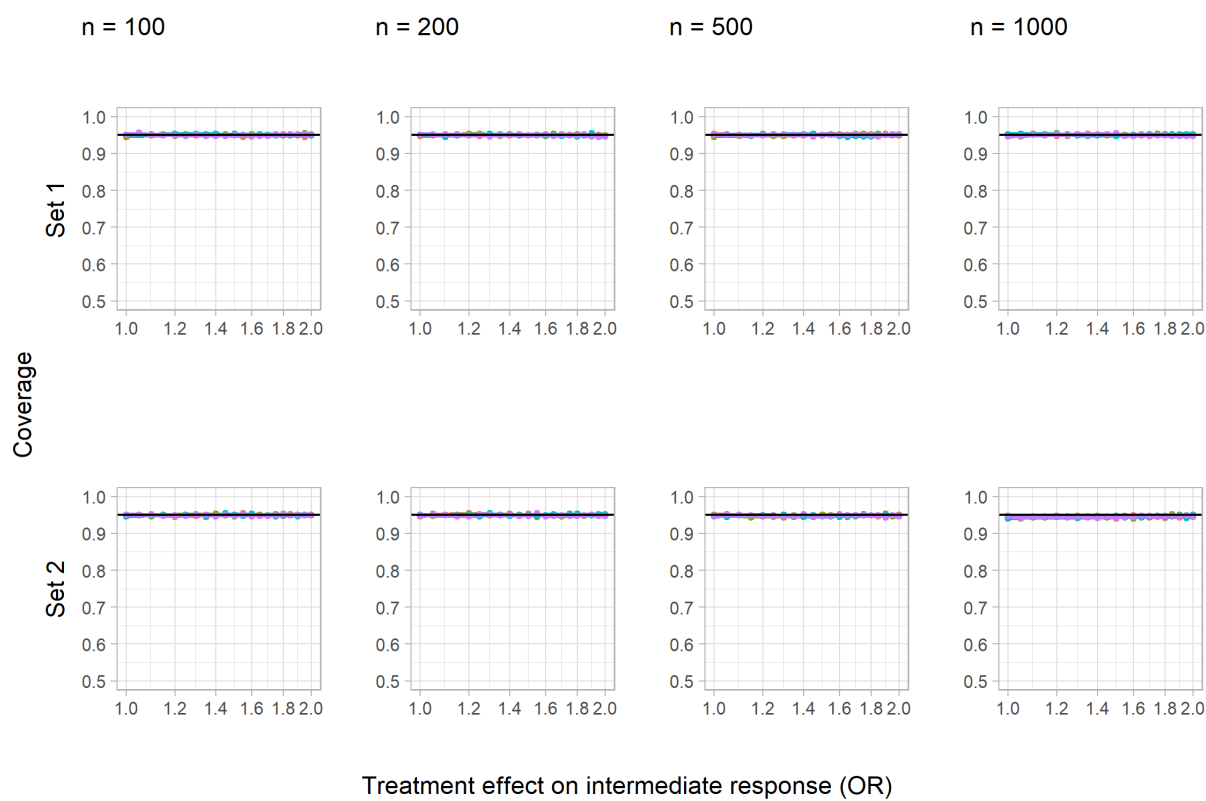

**Figure 3: Coverage of a 95% confidence interval corresponding to a simple difference in means in the continuous outcome simulation study (core scenarios). Colour indicates treatment effect on the outcome variable (SDs): red = 0, green = 0.2, blue = 1, purple = 5.**

Figure 3 shows that coverage was essentially at the nominal level in both core scenarios. This remained true in sensitivity analysis B, where the sign of the treatment effect on the intermediate was changed (S Figure 5), in sensitivity analysis C, where incidence of intermediate events was increased (S Figure 6) and for Set 1 in sensitivity analysis A, where confounding was increased (S Figure 4). However, in the increased confounding scenario, coverage was reduced for Set 2 with increasing sample size, and this was modified by the size of the treatment effect on the intermediate.

*Type 1 error*

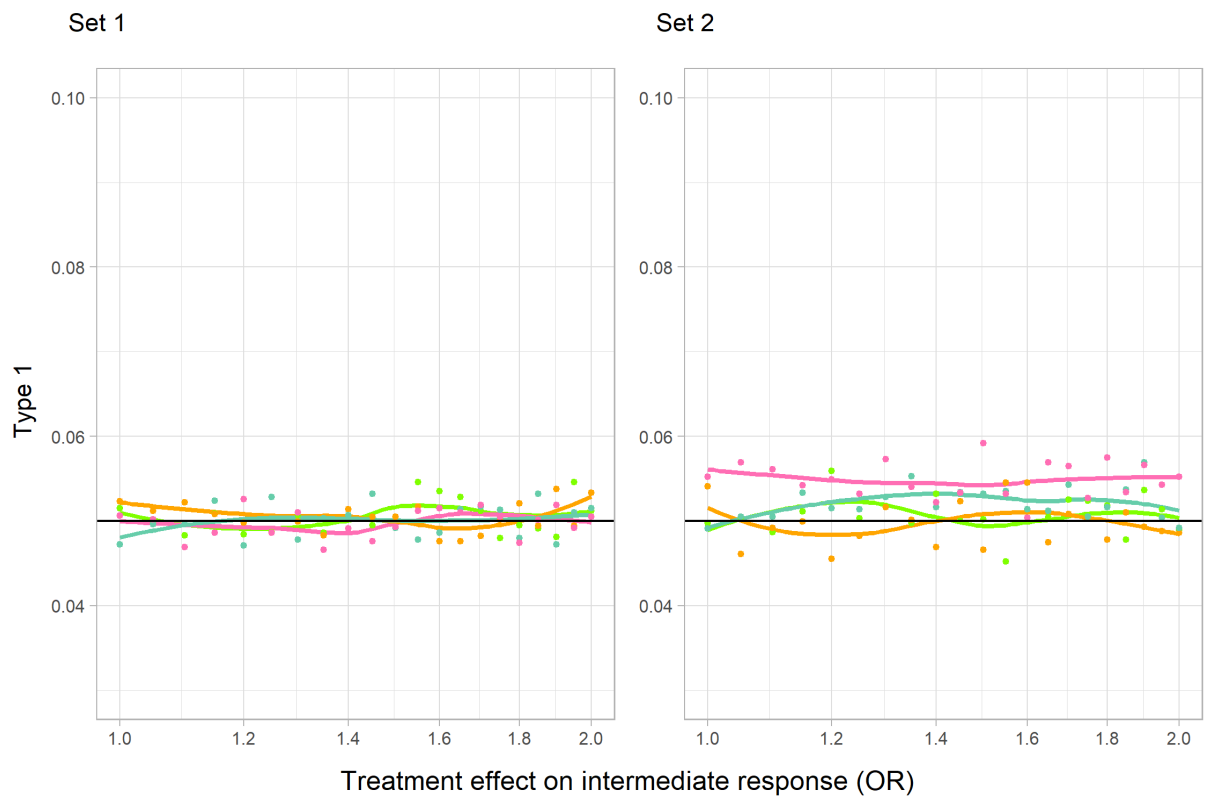

**Figure 4: Type 1 error of a t-test in the continuous outcome simulation study at a 5% significance level (core scenarios). Colour indicates the total starting sample size in each simulated trial (light green = 100, orange = 200, dark green = 500, pink = 1000). Horizontal line indicates the nominal level.**

In core scenarios, Type 1 error was not adversely affected in Set 1, but was noticeably increased with an interaction present (Set 2) and a large sample size (Figure 4). In sensitivity analysis A, with increased confounding, Type 1 error was inflated as treatment effect on the intermediate increased in the no interaction set (1), but remained close to the nominal value for lower, more plausible values (S Figure 7). By contrast, in Set 2 (interaction between treatment and $u_i$) Type 1 error was increased when the treatment effect on the intermediate was absent or small. The inflation increased with sample size, almost doubling for n = 1000. Sensitivity analyses B) and C) showed that neither changing the sign of the effect on the intermediate nor increasing the event rate substantially altered results compared to the core simulations – Type 1 error remained at the nominal level in Set 1, and became elevated at larger sample sizes in Set 2 (S Figures 8 and 9).

*Empirical and Model SE*

S Figures 10 to 17 show empirical and model SEs in the core and all sensitivity scenarios. Empirical and model SEs were similar (to each other) in any given scenario. SE decreased with increasing treatment effect on the intermediate variable, since this corresponded to more participants in the analysis subgroup overall (in the treatment arm). The impact of increasing treatment effect on precision was least when the intermediate event rate was increased (S Figures 13 and 17).

Binary outcome study

For smaller sample sizes (n = 100, 200) there were substantial amounts of missing data arising due to iterations where the treatment effect was inestimable (OR), because the tested scenarios frequently result in small numbers of participants with truncated outcome data, and so there are frequently zero outcome events in at least one arm (S Figure 18). Clearly, the proportion of missing data depends on the size of the treatment effect (and so is informative), which immediately suggests that the routine analysis of truncated endpoints in smaller (really, typical) trials might be problematic. The only sensitivity analysis for which this differed was C), which increased event rates, such that it was rare for no events to occur (S Figures 19-21).

The amount of missing data caused by an inability to calculate a test statistic was much lower than the amount caused by inability to estimate a treatment effect (excepting sensitivity analysis C)) but remained very high (in the region of 20%) for small sample sizes and modest/ realistic treatment effects, and was clearly related to the magnitude of the treatment effect on outcome (S Figures 22-25).

*Bias*

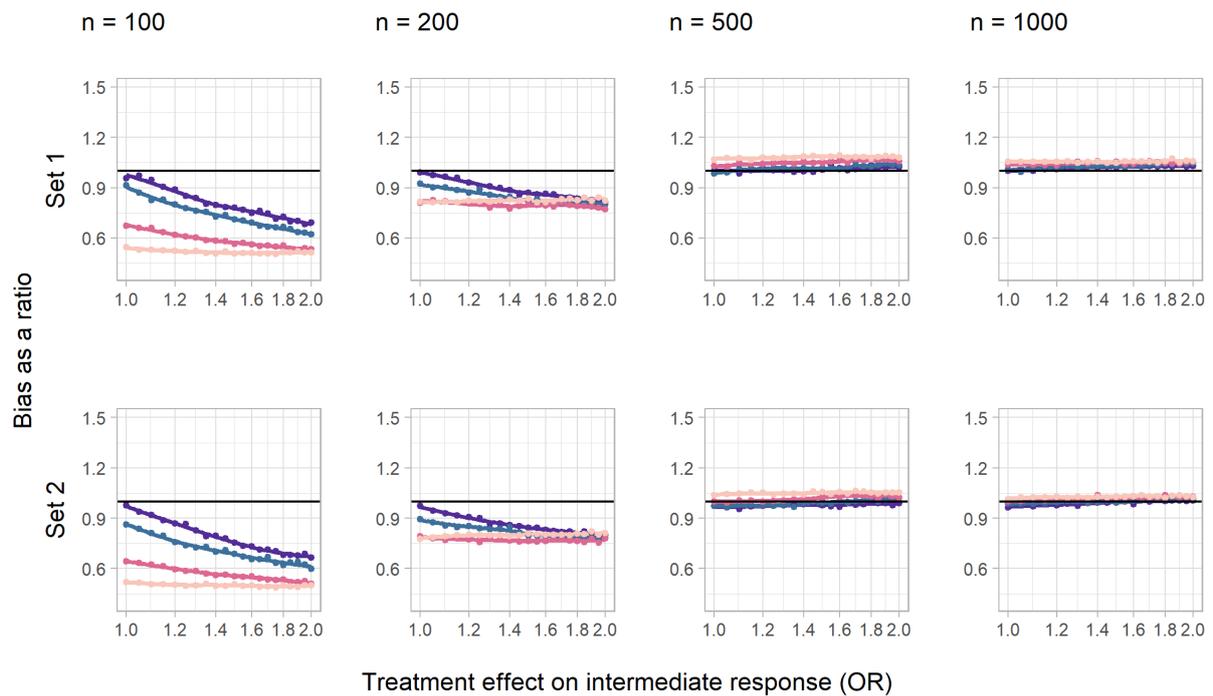

**Figure 5: Bias expressed as ratio of estimated to true odds ratio (ROR) in the binary outcome study (core scenarios). Colour indicates treatment effect on the outcome (ORs) (Purple = 1, blue = 1.2, darkpink = 2, light pink = 5)**

Bias, expressed as a ROR, fell below 1, substantially so for most of the tested treatment effect sizes, for sample sizes of 100 and 200 in both Sets (Figure 5). As noted above, the smaller sample sizes were subject to informative missing data, offering one explanation for the difference between the smaller and larger trials, for which the bias was much smaller, and in the opposite direction (ratios in the region of 1 to 1.05, or in the region of 1.08 when looking at the implausibly large treatment effect on the outcome, OR = 5). A comparison of the first two columns between S Figure 18 and Figure 5 actually show that missingness is negatively correlated with the size of the bias (the scenarios with less missing data have a ratio further from unity). However, sensitivity analysis B), where the effect of the treatment on the intermediate is reversed, shows that this is due to the fact that the influence of missing data (causing overestimation of the odds ratio) and of the treatment effect on the intermediate (causing underestimation of the odds ratio) act in opposite directions in the core scenarios (S Figure 27). Moreover, increasing the event rate, thereby eliminating the missing data issue, (sensitivity analysis C)) essentially removed the problem (S Figure 28). For trial sizes of 500 and 1000, increasing the treatment effect on the intermediate variable to an extreme value (OR of 5) did result in substantial bias even in the absence of interactions; ratios around 1.35 and 1.2 were observed for sample sizes of 500 and 1000, respectively. Increasing confounding in sensitivity analysis A) modified the relationship between treatment effect on the intermediate variable and bias for the larger sample sizes (steeper slope in S Figure 26 compared to Figure 5), although the bias remained modest for plausible parameter values.

*Coverage and SE*

The coverage level was too high at all effect sizes for smaller trials (Figure 6). This was at least partially attributable to missing data; the model SE was greater than the empirical SE of the computable estimates in the presence of missing data (S Figures 32 and 36). This was true in sensitivity analyses A) and B) (S Figures 29 and 30), but not C) where rates of missingness were minimised due to an increased event rate. For larger trial sizes, estimated coverage levels were generally close to, if not identical to, the nominal level, although discrepancies of several percentage points were apparent for large negative treatment effects on the intermediate (sensitivity analysis B, S Figure 30).

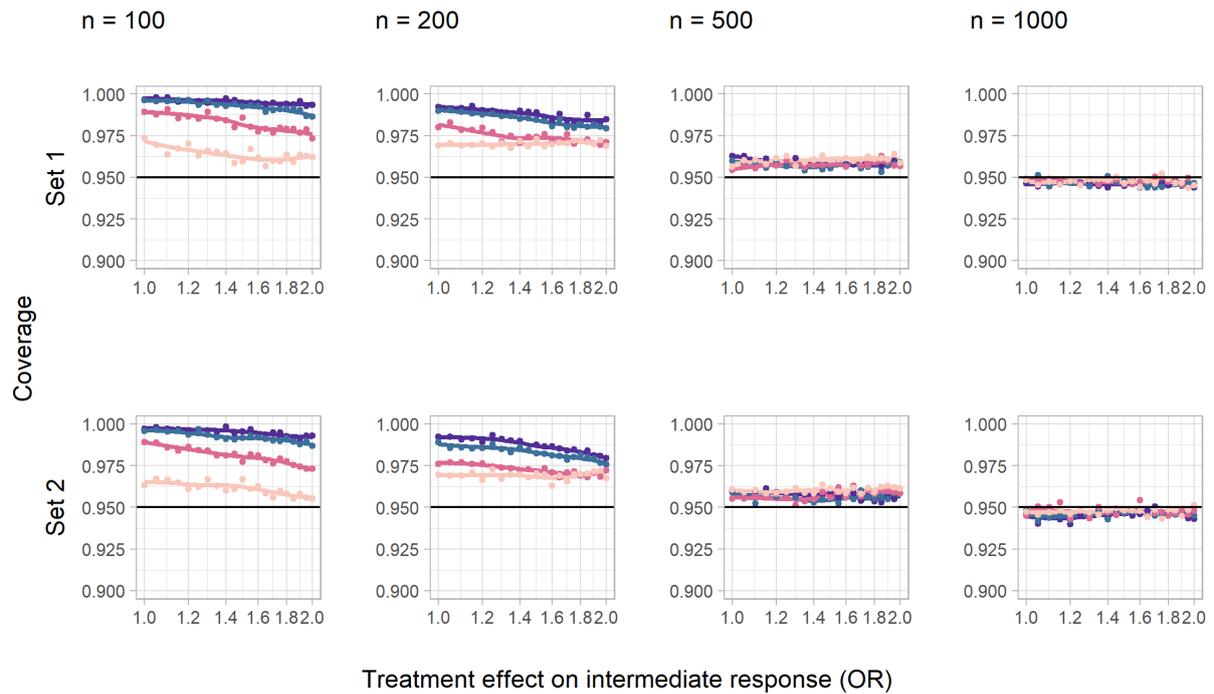

**Figure 6: Coverage of 95% confidence interval obtained using logistic regression in the binary outcome study (core scenarios). Colour indicates treatment effect on the outcome (ORs) (Purple = 1, blue = 1.2, darkpink = 2, light pink = 5).**

The rejection rate of both chi-squared tests were essentially equivalent, and Fisher's test performed consistently poorly. For trial sizes of 100 and 200, both subject to substantial amounts of missing data, Type 1 error fell below the nominal level for all three methods (Figure 7). The chi-squared tests achieved the nominal level for larger sample sizes in both Sets however, regardless of the size of treatment effect on the intermediate variable.

Increasing confounding (sensitivity analysis A)) and changing sign of the treatment effect on the intermediate (sensitivity analysis B)) did not change things – appropriate Type 1 error was observed for the chi-squared tests for larger trial sizes, but not for smaller trial sizes (S Figures 40 and 41). Increasing the event rate (sensitivity analysis C)) resulted in appropriate Type 1 error rates for chi-squared tests at all trial sizes, and an improvement in the performance of Fisher's test (S Figure 42) suggesting that issues in performance of the methods are primarily linked to informative missingness of calculated test statistics.

*Type 1 error of statistical tests*

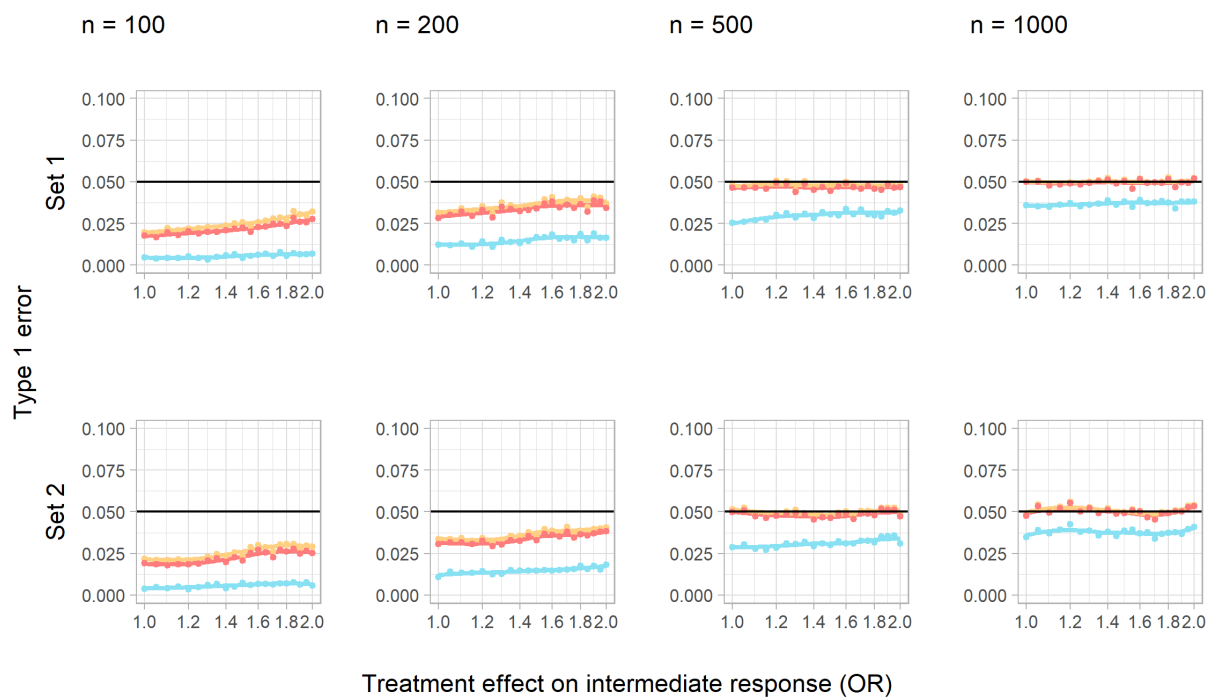

**Figure 7: Type 1 error of statistical tests in the binary outcome study (core scenarios). Fisher's exact test = blue, chi-squared test = pink, adjusted chi-squared test = yellow.**

**Discussion**

In the present study, we have created a simulation platform for studying the effects of outcome truncation in reproductive medicine trials. Using this platform, we present here a simulation study to characterise the phenomenon in scenarios resembling ART RCTs, although we believe the platform may also be of use for reproductive medicine trials beyond infertility, with similar structure (for example, the study of the effect of iron supplementation prior to pregnancy on perinatal outcomes (23), where treatment precedes the intermediate variable and outcomes are truncated).

We aimed to quantify the magnitude of the problems introduced by outcome truncation in practice, by using both comparative effectiveness and meta-epidemiological research in reproductive medicine to inform the simulation parameters. Our findings in this regard are therefore contingent upon the representativeness of the parameter values (24), and so we begin our discussion by briefly reviewing the motivations for the tested values and our confidence in these selections.

We opted to consider treatment effects on both the intermediate variable and the outcome variable that ranged from no effect up to implausibly large (an OR of 5 or difference in means of 5 SDs), in order to cover all bases with respect to these parameters. We note that we probably expect typical effects on the intermediate variable (for example, clinical pregnancy or live birth) to be less than ~1.2 when expressed as an OR. Typical treatment effects on live birth in Cochrane meta-analyses of infertility therapies were in the region of a few percentage points in a recent review (17), with larger estimates tending to arise from meta-analyses containing fewer study participants (generally representing less precise estimates). Moreover, these effect estimates are expected to be inflated to an unknown degree by publication bias. These considerations suggest that modest effects on the intermediate variables are to be expected. Typical effects on *outcome* variables subject to truncation are harder to establish, in part precisely because they are obscured by the censoring phenomenon under scrutiny in the present study. We have considered an expansive range of values for this parameter, and have found that its magnitude appears to relate to the performance of standard methods for analysis of truncated binary outcomes. Trialists using the platform to assist in the design of new studies are advised to consider a range of plausible values for this parameter, where plausibility is established in conjunction with clinical experts.

The strength of unmeasured confounding between the intermediate and outcome variables may turn out to be an important parameter. Increasing the strength of confounding in a sensitivity analysis modified the impact of increasing the treatment-on-intermediate effect, although neither the modification nor the consequences for performance were dramatic, for either continuous or binary outcomes. By contrast, an earlier study of pregnancy exposures and truncated continuous long-term health outcomes in children indicated substantial problems with naïve approaches to analysis (8), and our own attempts to replicate that study suggest that this was because the tested scenarios implied stronger confounding between the intermediate and outcome variable than we have considered here. If we have understated the plausible strength of confounding to a significant degree, we may have undersold the implications of outcome truncation. Speculating about the possible extent of unmeasured confounding between two variables is challenging, and in this case the correlation between the two cannot be directly observed (since the outcome is defined conditionally on the intermediate taking a particular value). Considering the total unexplained variation in each variable might be one way to start building intuition here, as might identification of known shared prognostic factors in the literature. This too is complicated by potential causal dependency between the intermediate and outcome variables however, as well as by the potential for distortion by the censoring phenomenon. In RCTs, it is likely that confounding will often be reduced through trial exclusion criteria. For example, restrictions are often placed on smoking and maternal BMI, which are associated with birthweight (25) (26). These restrictions will not eliminate confounding altogether however. For example, in the case of smoking, recent-quitters and never-smokers might still differ with respect to chances of conception and live birth in addition to birthweight. The potential for confounding might be strong in trials in populations with heterogeneous causes of infertility, where there might be underlying, unmeasured metabolic disease in some participants. Where knowledge of the other structural parameters is relatively strong, one use of the simulation platform presented here might be to investigate the strength of unmeasured confounding that would be required to introduce substantial bias or overturn a result.

A second function of the simulation study presented here was to elucidate the factors which affect performance of the standard analyses used in this context, and to the extent that we have allowed these factors to independently vary, these findings might not be so contingent on the particular parameter values used.

The present study is subject to other potential limitations. We have considered the situation where treatment is delivered on a single occasion, and the outcome is established at a single timepoint. These conditions are commonplace in ART studies, but studies evaluating cumulative outcomes over extended courses of treatment exist and are becoming more popular, since it is now recognised that outcomes after repeated attempts to conceive are particularly relevant to subfertile patients (27) (28). The present study is not directly relevant to these scenarios.

Another point to consider is that we have evaluated methods against a particular estimand, corresponding to the effect on outcome in the (hypothetical) absence of censoring. We selected this on the grounds that it aligned with a common interpretation given to ART trials subject to outcome truncation, e.g. (10), and that it frequently has clinical relevance in this context. Taking the effect of embryo culture medium on birthweight as an example, it would be useful for a clinician to know if a particular advantageous medium (in terms of live birth rate) resulted in reduced birthweights by adversely affecting foetal development, or else if reduced birthweights were an inevitable consequence of improving live birth rates in the population. This knowledge might influence the decision-making process undertaken by patients and clinicians. For instance, the potential harms to offspring might be considered unacceptable, or the availability of a co-intervention known to improve live birth rate might make an alternative less effective (live birth) but safer (birthweight) medium a more attractive choice. We would stress however that analysis of trial outcome data alone is unlikely to provide sufficient insight into this sort of mechanistic hypothesis, and must be considered alongside biological evidence.

Other estimands have been described in the context of competing risks however (29, 30), and each of these might be more or less attractive depending on the particulars of the research question under evaluation and the assumptions the study team are willing to make. Proposals include (what has been described as) a *total effect* of treatment, in which a composite outcome is defined for everyone regardless of their status with respect to the intermediate variable; anyone not experiencing the intermediate event is classified as not having the outcome (29, 31) see also the *treatment policy strategy* in (18, 32). For example, any participants who did not become pregnant would be considered not to have had a miscarriage. Under this definition, a treatment could reduce the miscarriage rate by reducing the pregnancy rate, which does not conform to any intuitive notion of therapeutic benefit. Furthermore, this definition cannot be extended to continuous outcomes. Another potential hypothetical estimand is the Survivor Average Causal Effect, the effect in patients who would have had the intermediate event under either treatment allocation, (6) which raises questions about relevance to real patients. Another proposal is to consider direct and indirect separable effects, which require the analyst to postulate distinct causal pathways including and excluding the intermediate (30). This requires the intermediate variable to be construed as a mediator. We largely agree with the commentary of Snowden and colleagues (33) however, which clarifies the role of the intermediate variable in this context; the intermediate does not mediate the effect of treatment, so much as determine whether the outcome variable is defined. In light of the conceptual difficulty of interpreting the intermediate as a mediator, we have not considered a causal path from the intermediate to the outcome in the present study, but have included the option to do so in the simulation platform we provide. We have also not considered the potential role of an interaction between treatment effect on the outcome (rather than on the intermediate) and confounding factors here. We include the option to do so in the simulation code, but urge the user to consider whether an alternative estimand might be more appropriate in the presence of such an interaction.

With these considerations in place, we turn to the findings of the simulation study. In the continuous outcome study, the impact of outcome truncation on simple analyses based on the observed difference in means was less severe than had perhaps been anticipated, with reasonable bias, coverage, and Type 1 error rates for more realistic treatment effects, except in the scenario with increased confounding, when performance was notably affected in the presence of an interaction between treatment and the unmeasured confounder. These results might therefore be seen as relatively reassuring in relation to continuous outcome measures, depending on the plausible extent of unmeasured confounding and scope for interaction effects for the particular research question at hand. In particular, these results appear supportive of the finding in (10), that choice of embryo culture medium can influence birthweight of offspring born from ART.

The situation with binary outcomes appears somewhat more nuanced. For larger trial sizes of 500 or 1000, bias of the odds ratio was present but was relatively modest, and coverage was close to the nominal level, again provided that no interaction between treatment and the intermediate was present. These findings held when we increased the level of confounding and when we changed the direction of the treatment effect on the intermediate variable, to rule out the possibility of effects in opposing directions concealing problems. For these larger trial sizes then, our results appear to be qualitatively concordant with the conclusions of previous authors e.g. (3, 14), at least in the sense that bias was caused by outcome truncation, and was affected by increasing treatment effect on the intermediate variable, strength of unmeasured confounding and presence of interactions. Quantitatively however, we find that, within the parameters considered here, outcome truncation might not be so great a cause for concern (at least for large trials) as has previously been suggested. Indeed, close inspection of previous simulation results suggests that substantive performance issues have been observed only under parameter settings that would be quite extreme in the context of ART trials, e.g. large effects of exposure on intermediate (3). The Type 1 error rates for two variants of a chi-squared test were also close to the nominal level for larger trial sizes. Fisher's test performed poorly in this context, which may be attributable to the violation of the assumption of fixed margins, and this was presumably caused by varying numbers of participants entering the analysis set across simulated datasets within any given scenario.

For smaller trial sizes (n = 100, 200) outcome truncation creates serious challenges for the study of binary outcomes, and this appears to be attributable to separation (studies in which all or none of the analysable participants in a study arm have the outcome event). The current study highlights the fact that the likelihood of obtaining an effect estimate is related to the effect of the treatment on both the intermediate and outcome variables. As such, the subset of studies in which an effect estimate is calculable will not produce an unbiased sample for the purpose of estimation. Notably, small studies in ART are commonplace (17). Although small studies are unlikely to use a truncated response variable as a primary outcome, they may still be reported as secondary outcomes. There may be implications for systematic reviews here, since truncated binary secondary (as opposed to primary) outcomes, analysed in the post-randomisation subgroup, often appear in meta-analysis (examples). By design, meta-analyses incorporate all studies, including the smaller ones. Pooled estimates are therefore likely to be based on an informative selection process, leading to bias. This situation is subject to additional complexities compared to the usual case of meta-analysis of sparse events, and appears to warrant further investigation. In the interim, it is recommended to follow the advice set out in the Cochrane Handbook, which is to avoid meta-analysis of truncated outcomes wherever possible (5).

Comparisons of outcomes in the subgroup, e.g. (34), cited in (33) have been endorsed on the grounds that this represents the population at risk. A problem with this proposal in principle is that the observed difference in a trial, not representing an effect of treatment *per se*, will not apply wherever confounding or the effect on the intermediate (for example, due to differences in other aspects of the ART treatment protocol) differ. It may however be a useful quantity to consider from a public health perspective, provided it is based on a representative sample, and the present study suggests that simple analyses will yield reasonable answers in many cases. Nonetheless, it remains reasonable to seek alternative analytic methods that will be robust to outcome truncation under a broader range of data generating models. We note that unadjusted analyses, as are commonly performed in the field and as considered here, are unlikely to be optimal regardless of outcome truncation, since adjusting for prognostic covariates in a trial will improve precision (35-37). To the extent that the adjustment variables coincide with the confounders of the intermediate-outcome relationship, it is possible that performance might be improved compared to the unadjusted approach in the truncated outcome scenario, although with small samples and binary outcomes, it is possible that covariate adjustment might exacerbate issues relating to sparse data and separation (38). It remains to examine this empirically, as well as the most suitable approach for adjustment (e.g. regression versus inverse probability weighting approaches) and the implications for interpretation. In the presence of separation, Firth's logistic regression correction (39) has been recommended (40-42), and the performance of this approach for truncated binary outcome data warrants investigation. Methods to estimate the survivor average causal effect have been described (6, 7), as have sensitivity analyses designed for this context (43-45). Another proposal would be to consider a joint model of the intermediate and outcome variables, although it is not clear that this would be estimable in a point treatment setting. Methods for meta-analysis of truncated outcomes with small studies appears to be another avenue for future research.

**Conclusions**

In general, proposed approaches to analysis require substantial assumptions and relevant data (notably on sufficient confounding sets) to restore unbiased effects. Our simulation platform provides a rapid assessment of the implications of outcome truncation given user-input parameters, and can be used to assist in the design and interpretation of reproductive medicine trials, particularly in the case of small trials for binary primary outcomes or where there are expected to be strong confounders related to selection (conception or live birth) or interactions with treatment thereof. Finally, since the code is freely available for modification, we hope that it may serve as a platform for future methodological research in the area.

**Declarations**

*Ethics approval and consent to participate*

This simulation study did not require ethical approval.

*Consent for publication*

Not applicable.

*Availability of data and materials*

Code to reproduce this study is available at https://osf.io/gzqbr/ .

*Competing interests*

JW declares that publishing research benefits his career.

*Funding*

JW is supported by a Wellcome Institutional Strategic Support Fund award [204796/Z/16/Z].

*Author contribution*

JW conceived the idea, wrote the code and performed the simulation study. All authors contributed to the design of the study and interpretation of results. Additionally, all authors contributed intellectual content, wrote the manuscript, and approved the final version.

# Supplementary Figures

## Contents

# Continuous outcome study

## Bias

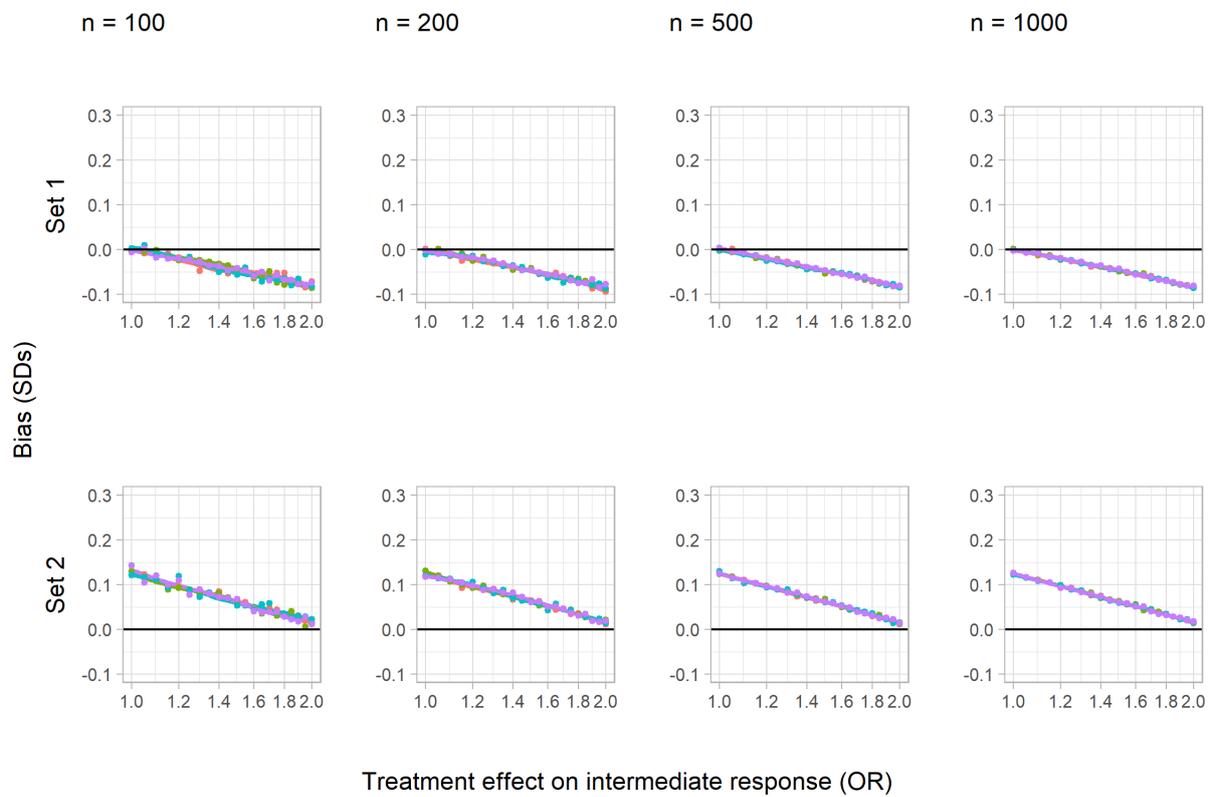

S Figure 1: Bias of a simple difference in means in the continuous outcome simulation study (sensitivity analysis A, increased confounding). Colour indicates treatment effect on the outcome variable (SDs): red = 0, green = 0.2, blue = 1, purple = 5.

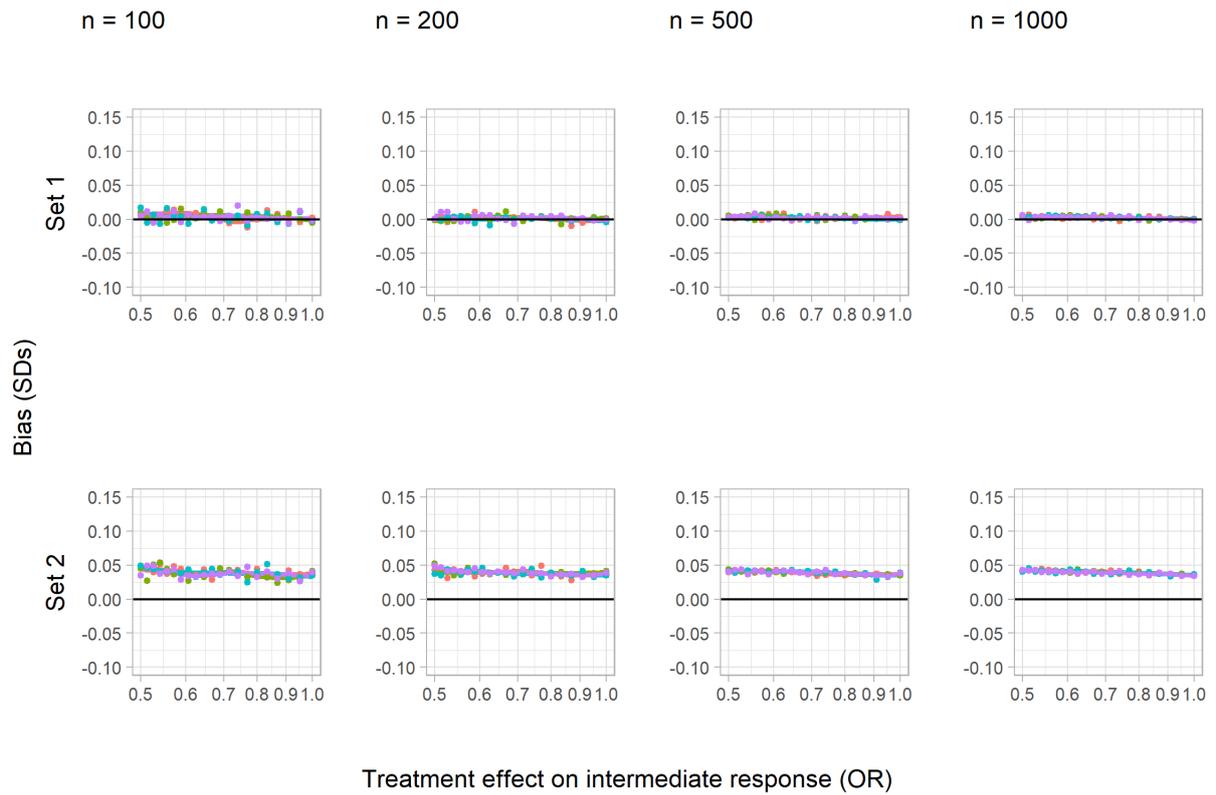

S Figure 2: Bias of a simple difference in means in the continuous outcome simulation study (sensitivity analysis B, changed direction of treatment effect on intermediate). Colour indicates treatment effect on the outcome variable (SDs): red = 0, green = 0.2, blue = 1, purple = 5.

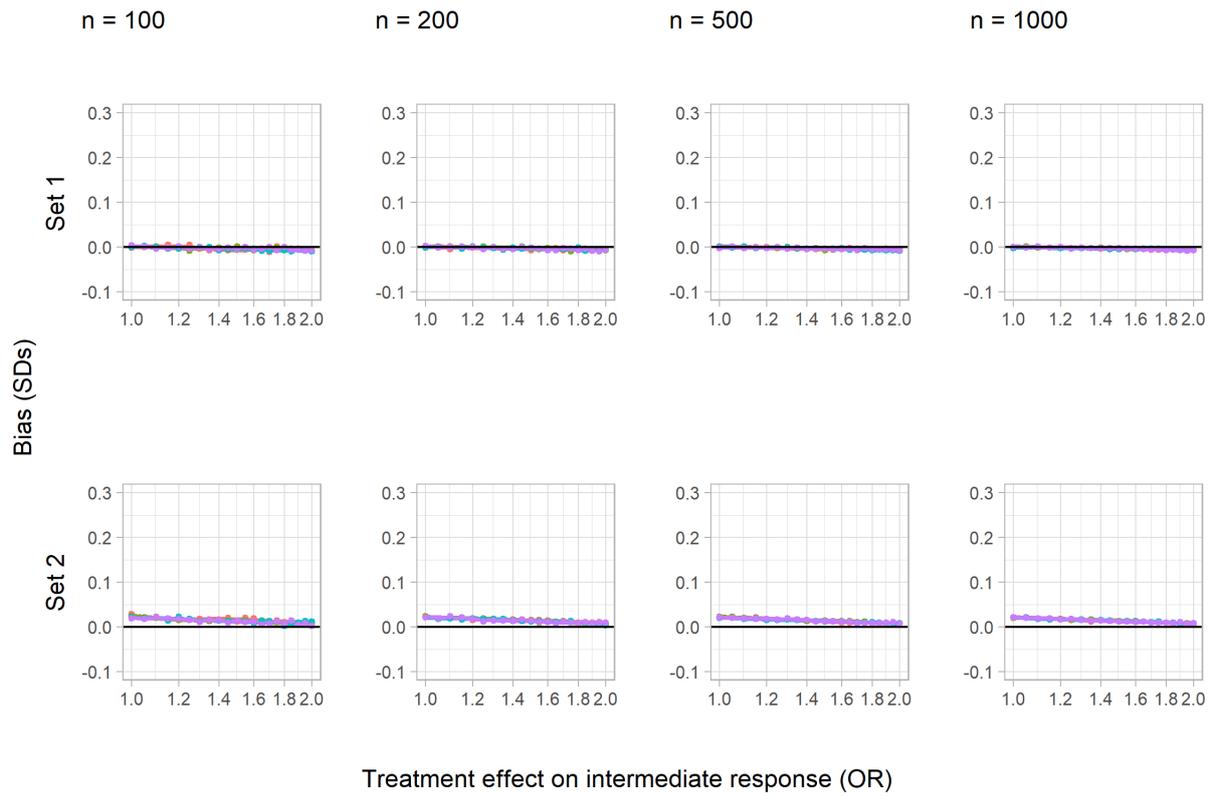

*S Figure 3: Bias of a simple difference in means in the continuous outcome simulation study (sensitivity analysis C, increased event rate). Colour indicates treatment effect on the outcome variable (SDs): red = 0, green = 0.2, blue = 1, purple = 5.*

# Coverage

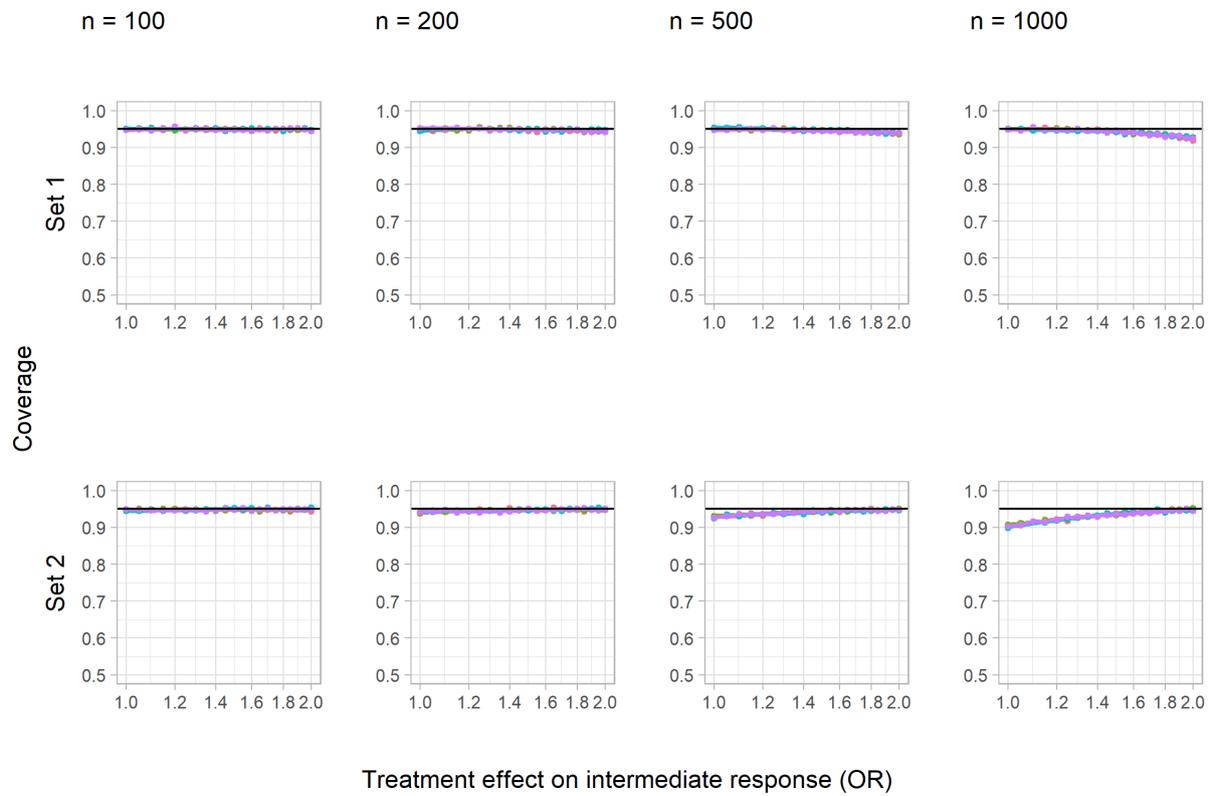

S Figure 4: Coverage of a 95% confidence interval corresponding to a simple difference in means in the continuous outcome simulation study (sensitivity analysis A), increased confounding). Colour indicates treatment effect on the outcome variable (SDs): red = 0, green = 0.2, blue = 1, purple = 5.

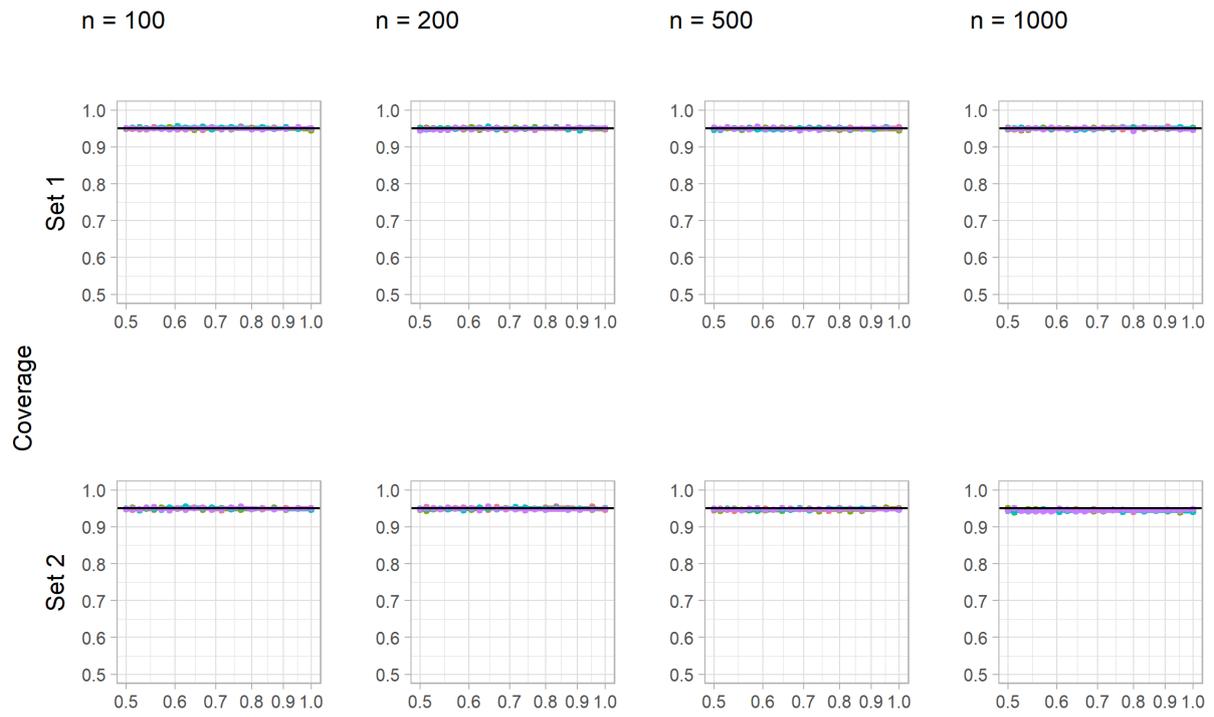

*S Figure 5: Coverage of a 95% confidence interval corresponding to a simple difference in means in the continuous outcome simulation study (sensitivity analysis B) changed direction of treatment effect on intermediate). Colour indicates treatment effect on the outcome variable (SDs): red = 0, green = 0.2, blue = 1, purple = 5.*

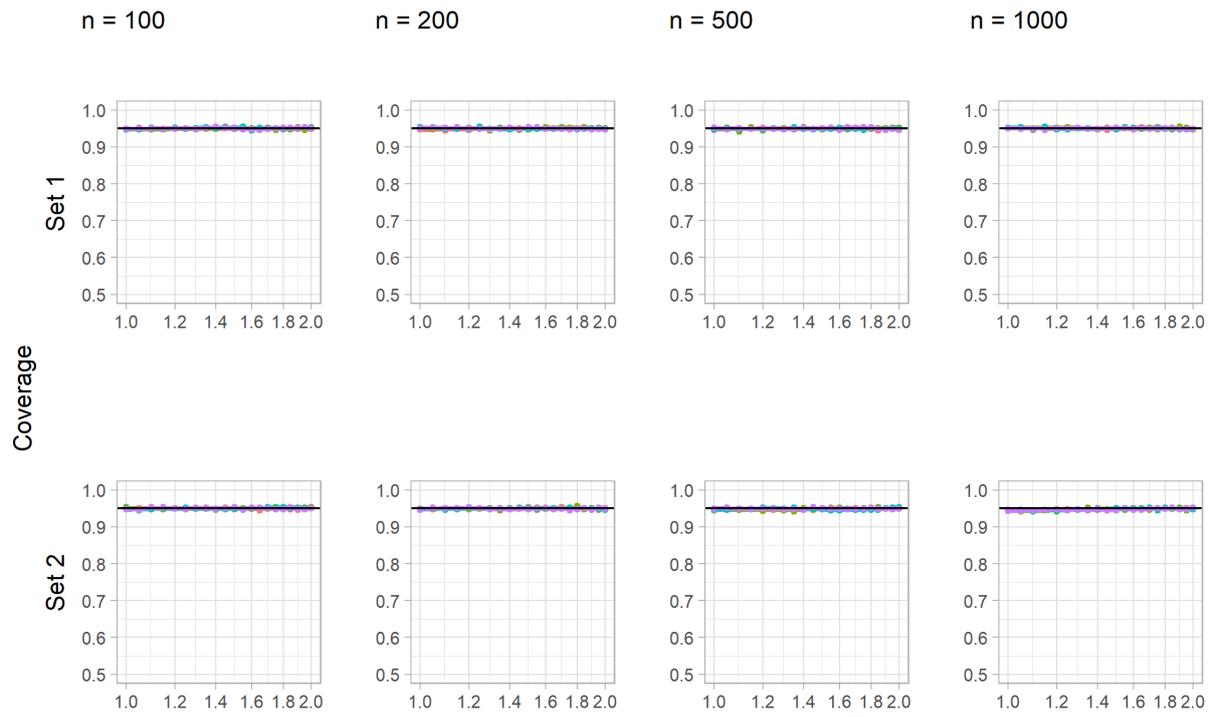

S Figure 6: Coverage of a 95% confidence interval corresponding to a simple difference in means in the continuous outcome simulation study (sensitivity analysis C), increase event rate). Colour indicates treatment effect on the outcome variable (SDs): red = 0, green = 0.2, blue = 1, purple = 5.

## Type 1 error

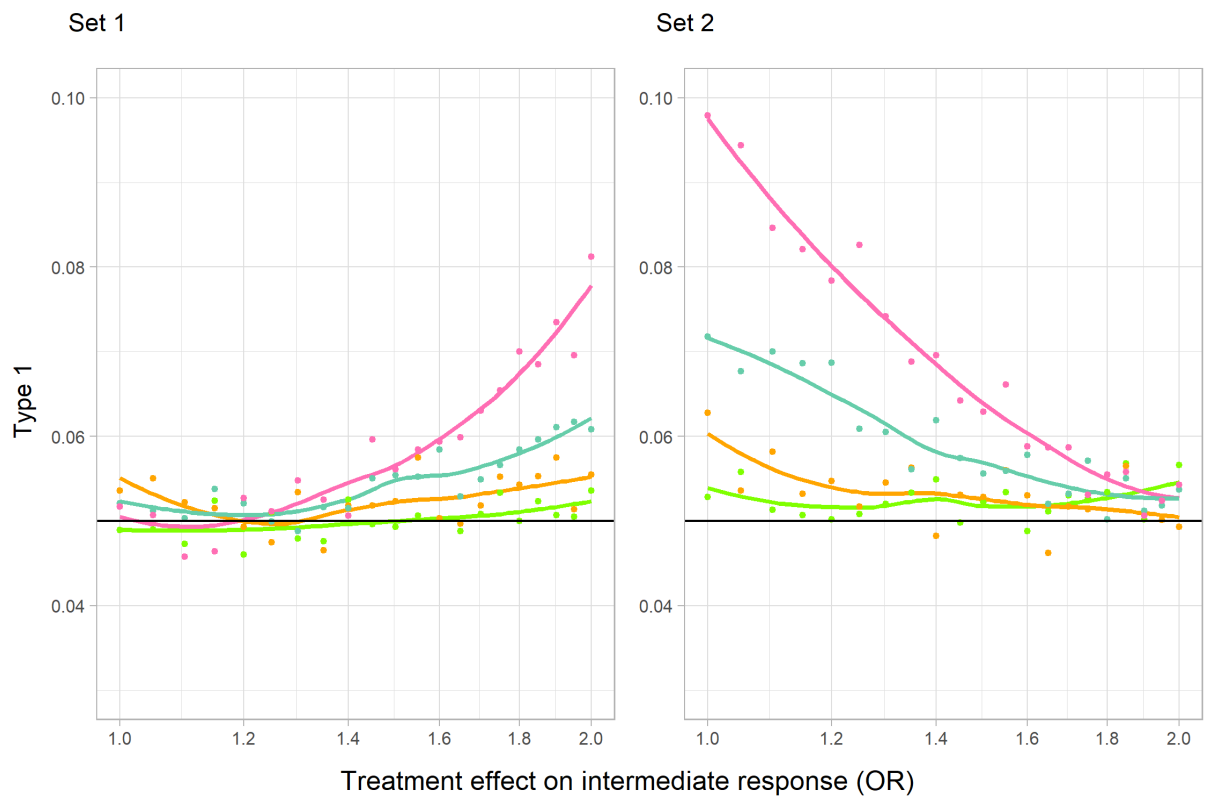

*S Figure 7:* Type 1 error of a t-test in the continuous outcome simulation study at a 5% significance level (sensitivity analysis A), increased confounding). Colour indicates the total starting sample size in each simulated trial (light green = 100, orange = 200, dark green = 500, pink = 1000). Horizontal line indicates the nominal level.

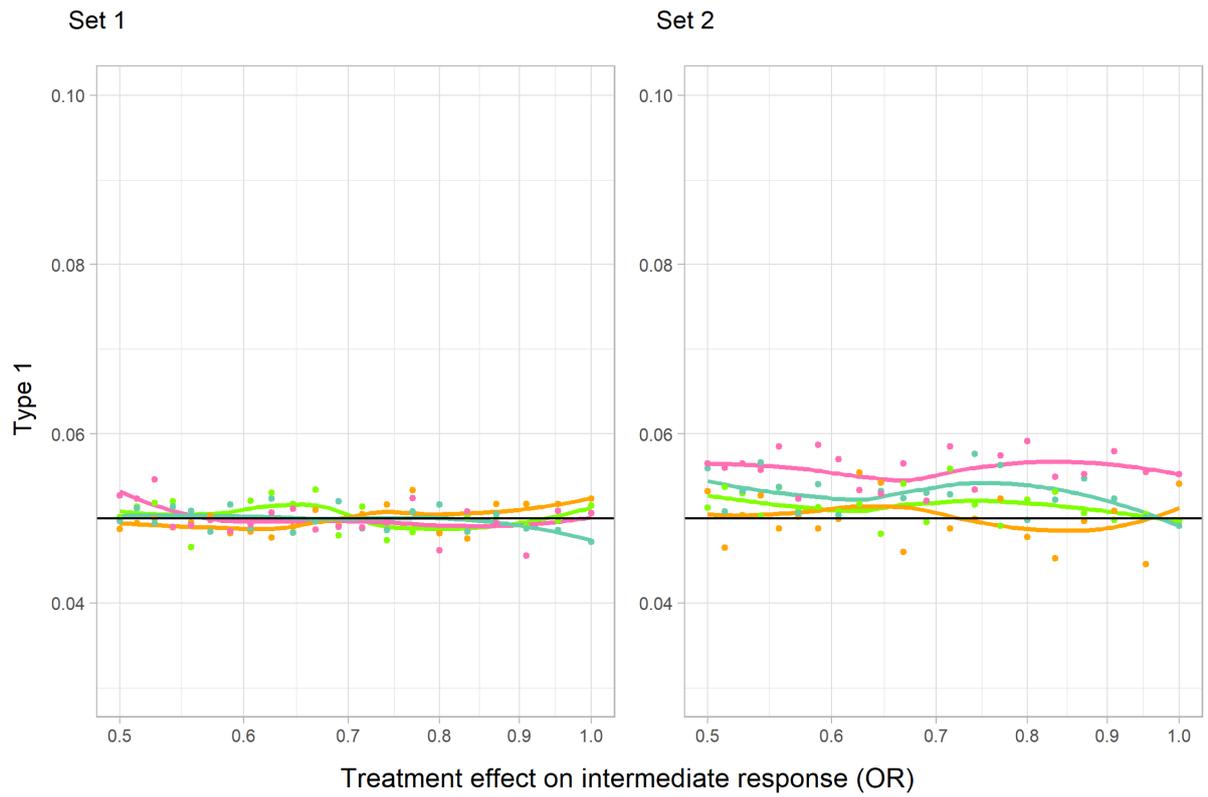

S Figure 8: Type 1 error of a t-test in the continuous outcome simulation study at a 5% significance level (sensitivity analysis B) changed direction of effect on intermediate). Colour indicates the total starting sample size in each simulated trial (light green = 100, orange = 200, dark green = 500, pink = 1000). Horizontal line indicates the nominal level.

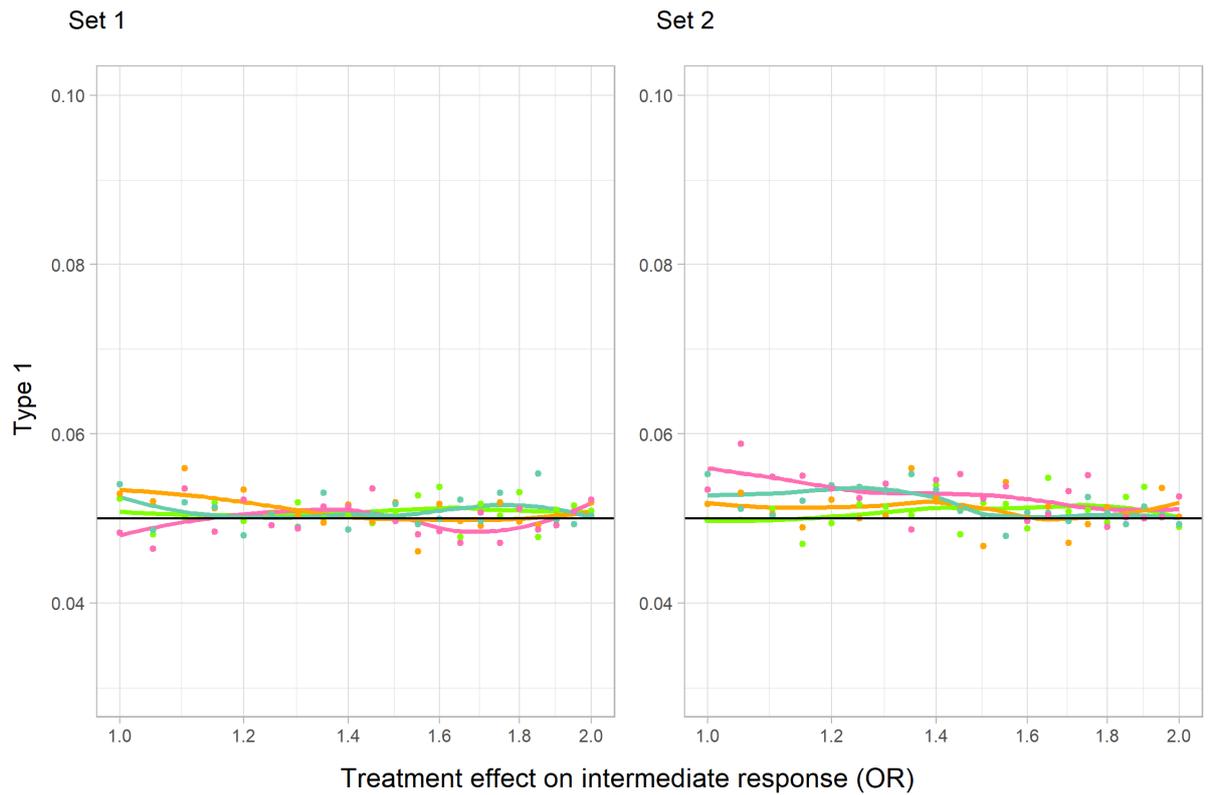

*S Figure 9:* Type 1 error of a t-test in the continuous outcome simulation study at a 5% significance level (sensitivity analysis C), increased event rate). Colour indicates the total starting sample size in each simulated trial (light green = 100, orange = 200, dark green = 500, pink = 1000). Horizontal line indicates the nominal level.

Empirical SE

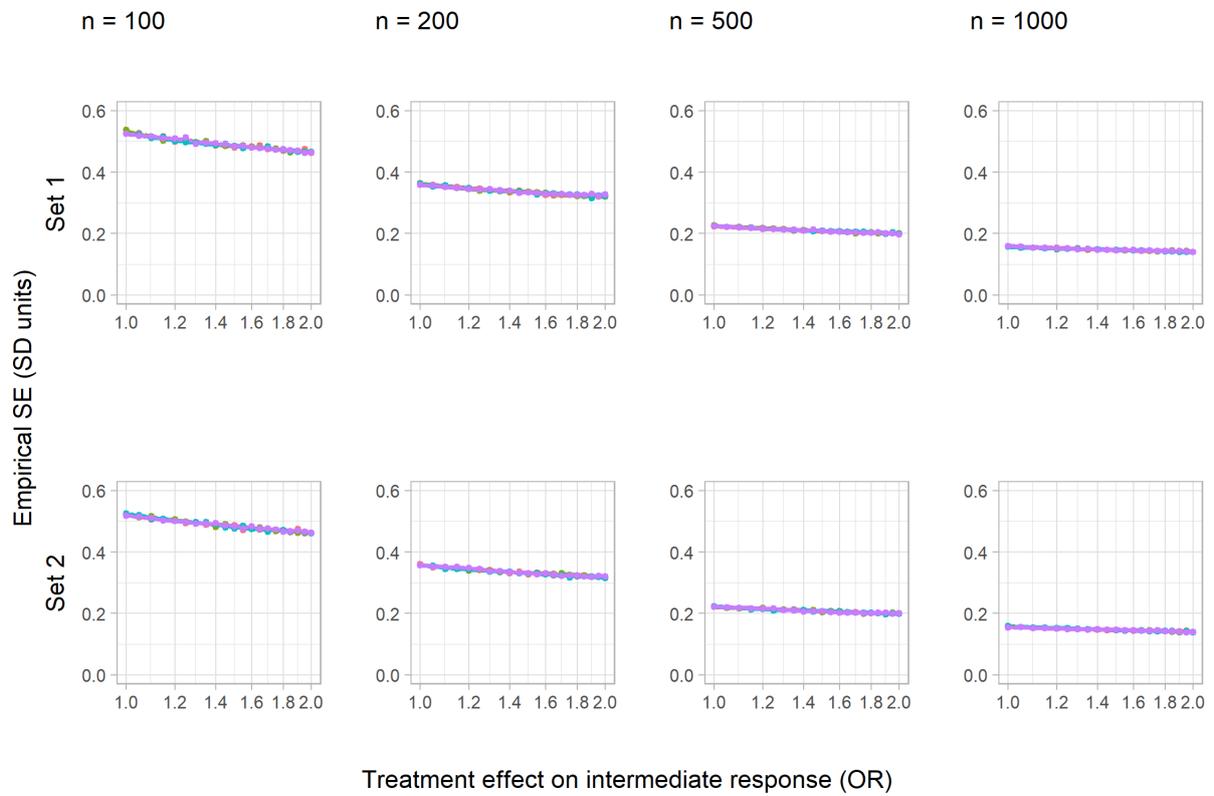

*S Figure 10: Empirical SE of a simple difference in means in the continuous outcome study (core senarios). Colour indicates treatment effect on the outcome variable (SDs): red = 0, green = 0.2, blue = 1, purple = 5.*

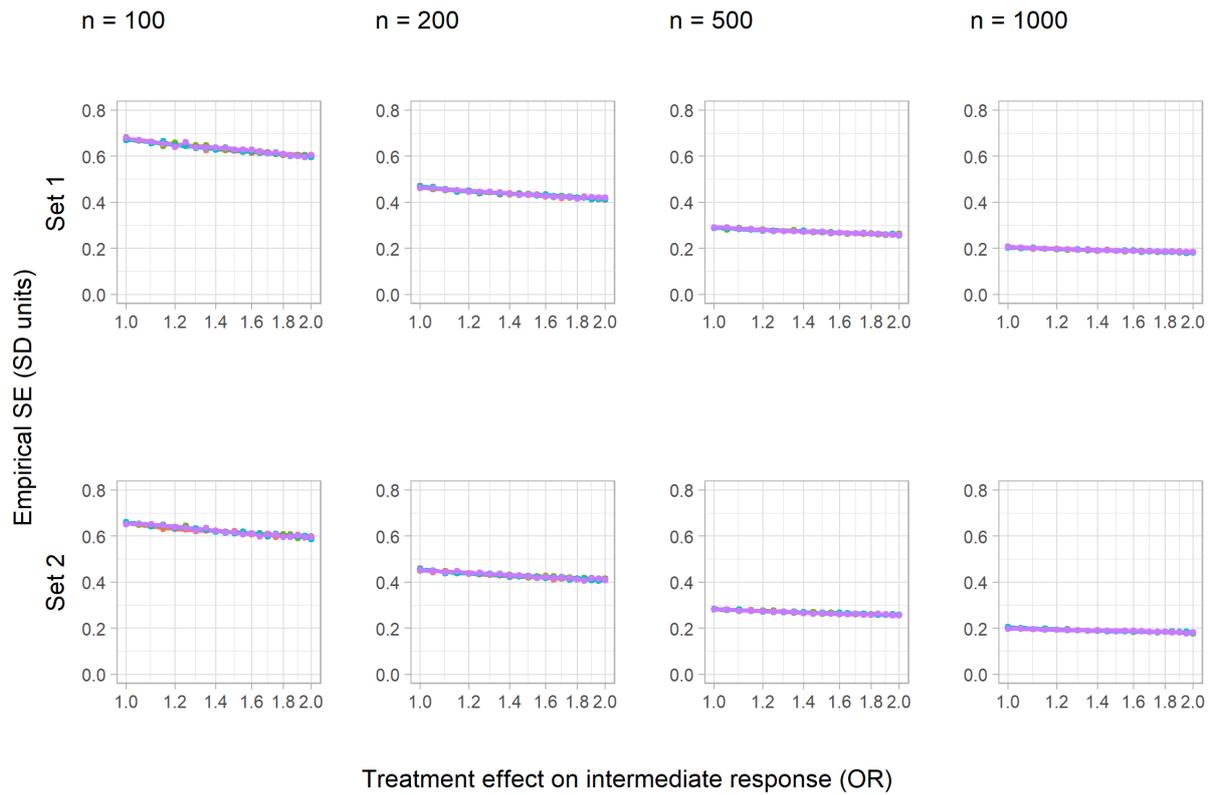

*S Figure 11: Empirical SE of a simple difference in means in the continuous outcome study (sensitivity analysis A), increased confounding). Colour indicates treatment effect on the outcome variable (SDs): red = 0, green = 0.2, blue = 1, purple = 5.*

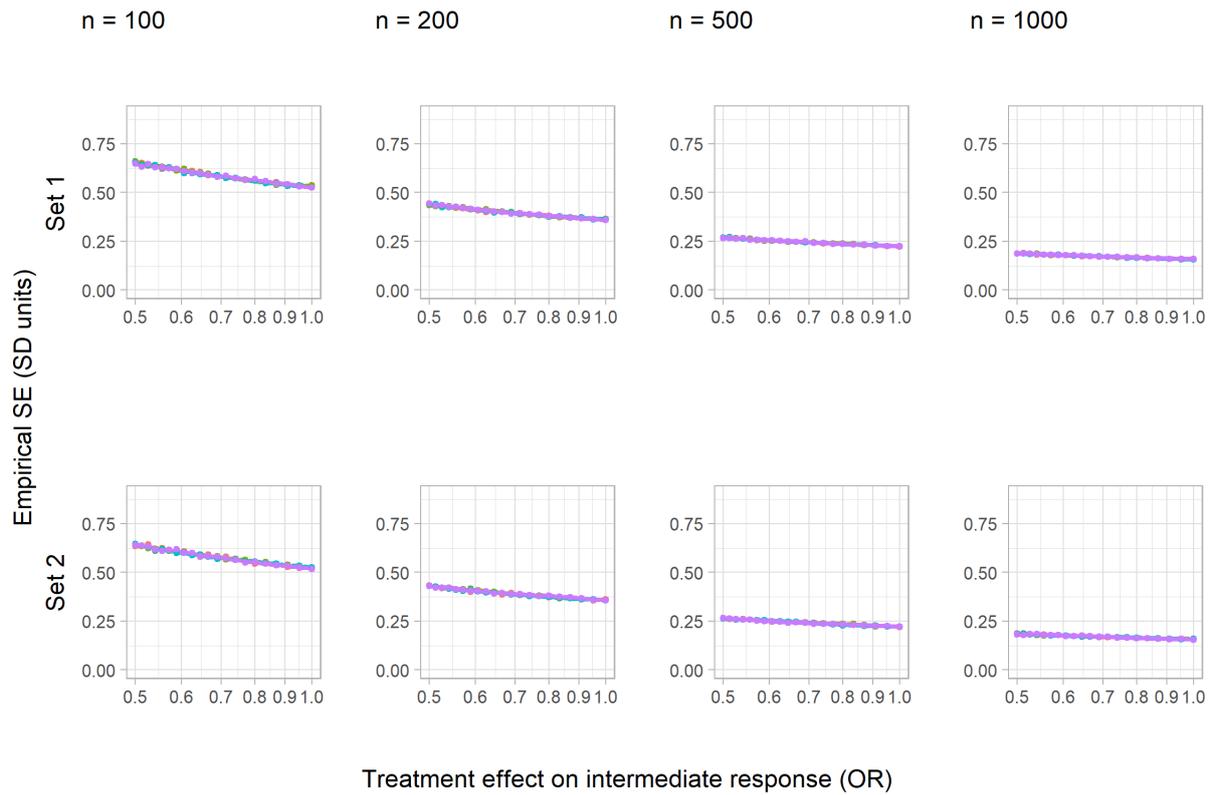

*S Figure 12: Empirical SE of a simple difference in means in the continuous outcome study (sensitivity analysis B), changed direction of treatment effect on intermediate). Colour indicates treatment effect on the outcome variable (SDs): red = 0, green = 0.2, blue = 1, purple = 5.*

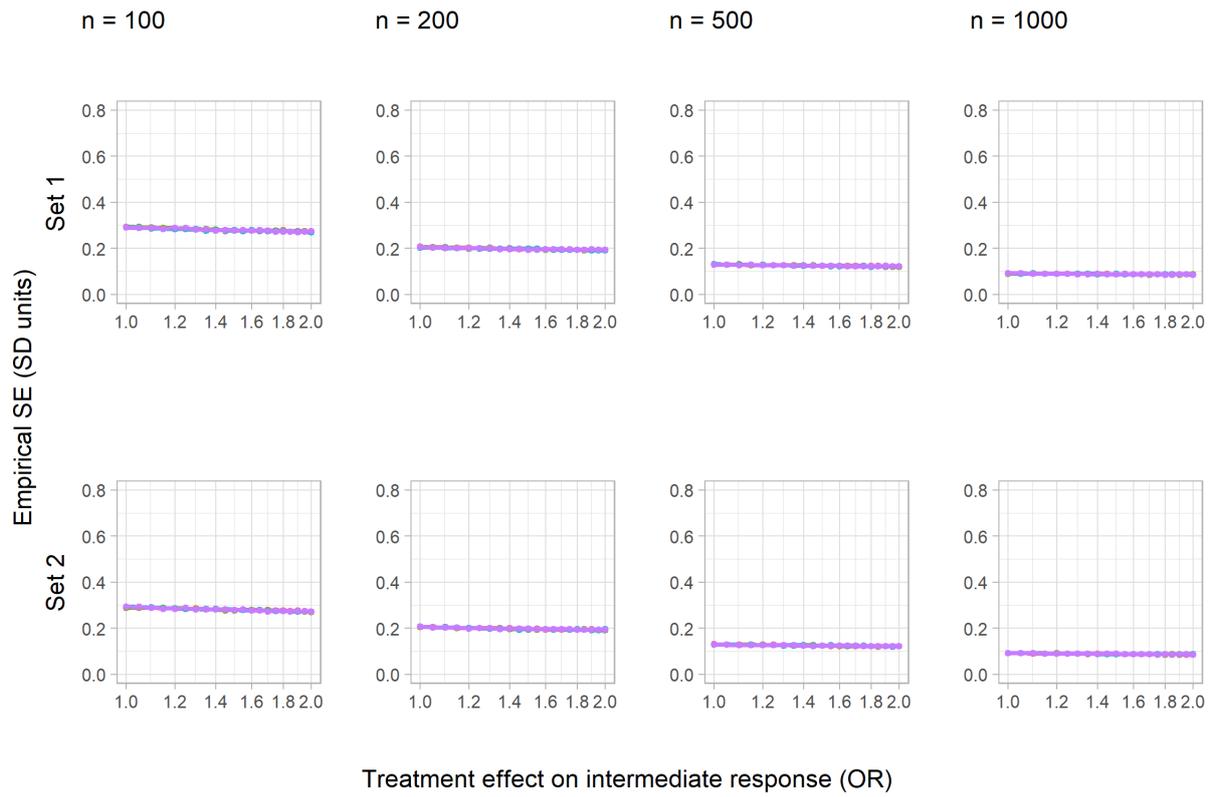

*S Figure 13: Empirical SE of a simple difference in means in the continuous outcome study (sensitivity analysis C), increased event rate). Colour indicates treatment effect on the outcome variable (SDs): red = 0, green = 0.2, blue = 1, purple = 5.*

## Model SE

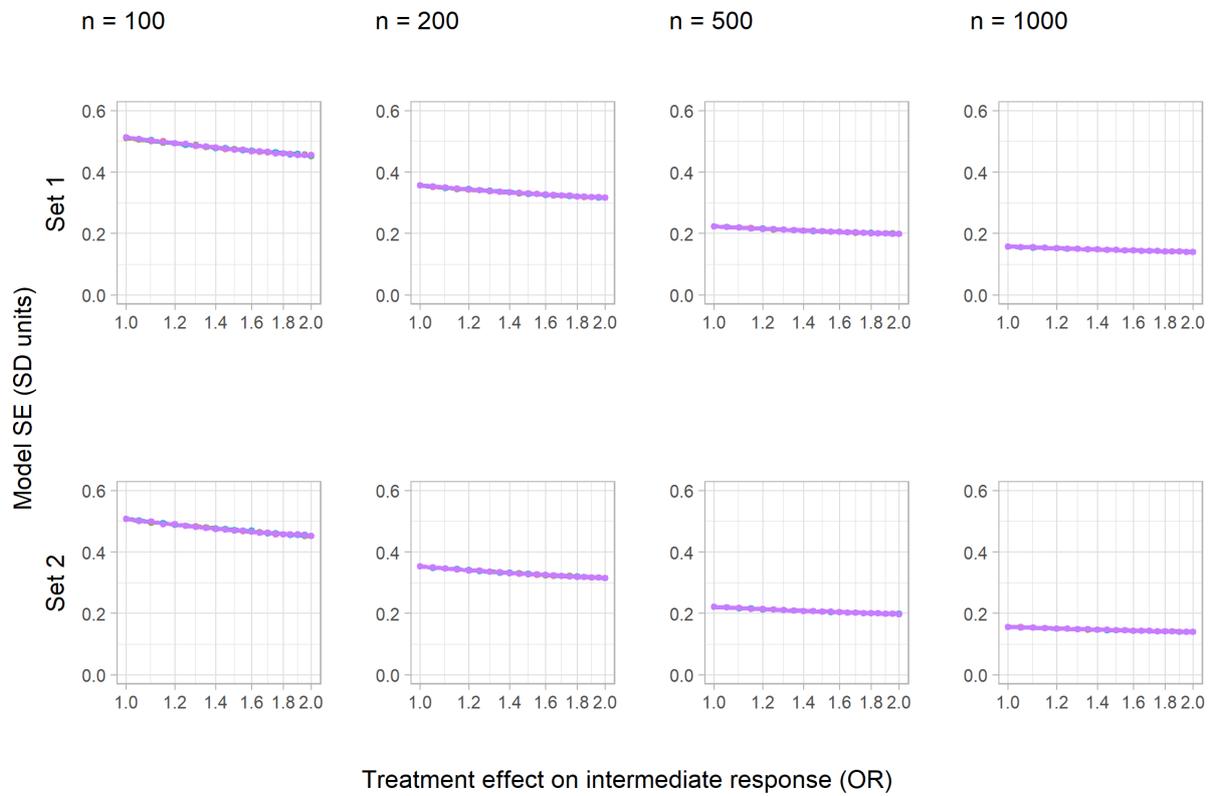

S Figure 14: Model SE of a simple difference in means in the continuous outcome study (core scenarios). Colour indicates treatment effect on the outcome variable (SDs): red = 0, green = 0.2, blue = 1, purple = 5.

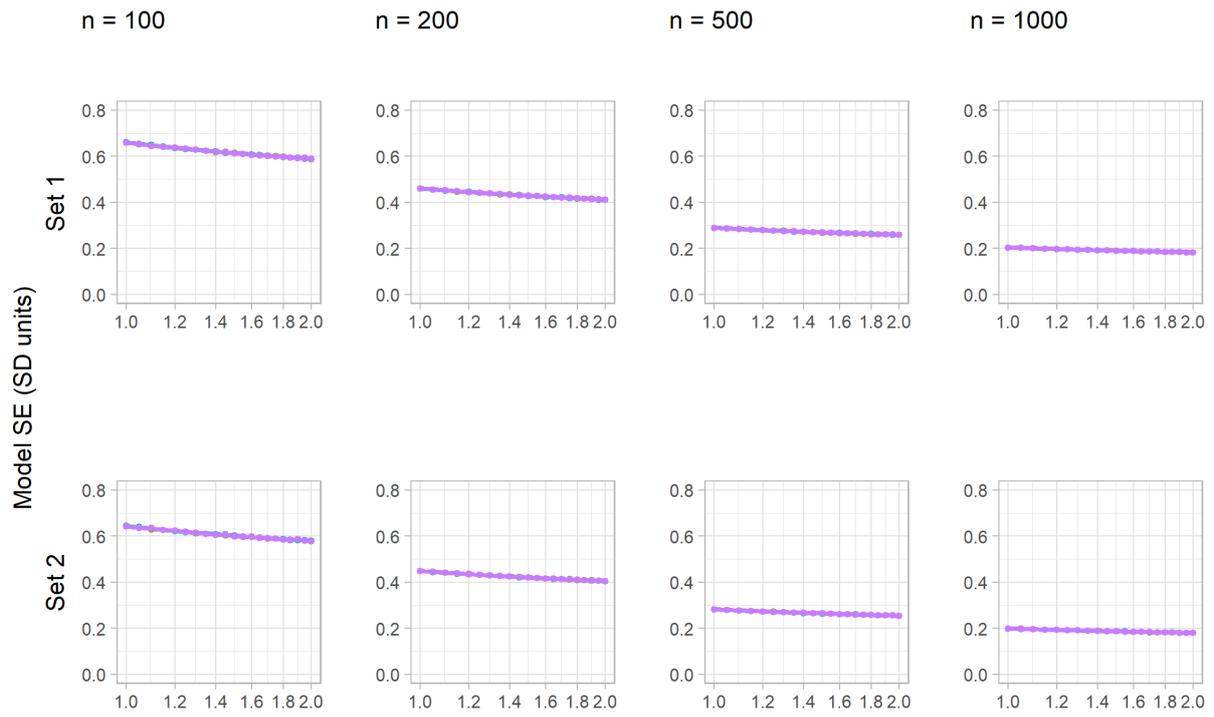

*S Figure 15: Model SE of a simple difference in means in the continuous outcome study (sensitivity analysis A) increased confounding). Colour indicates treatment effect on the outcome variable (SDs): red = 0, green = 0.2, blue = 1, purple = 5*

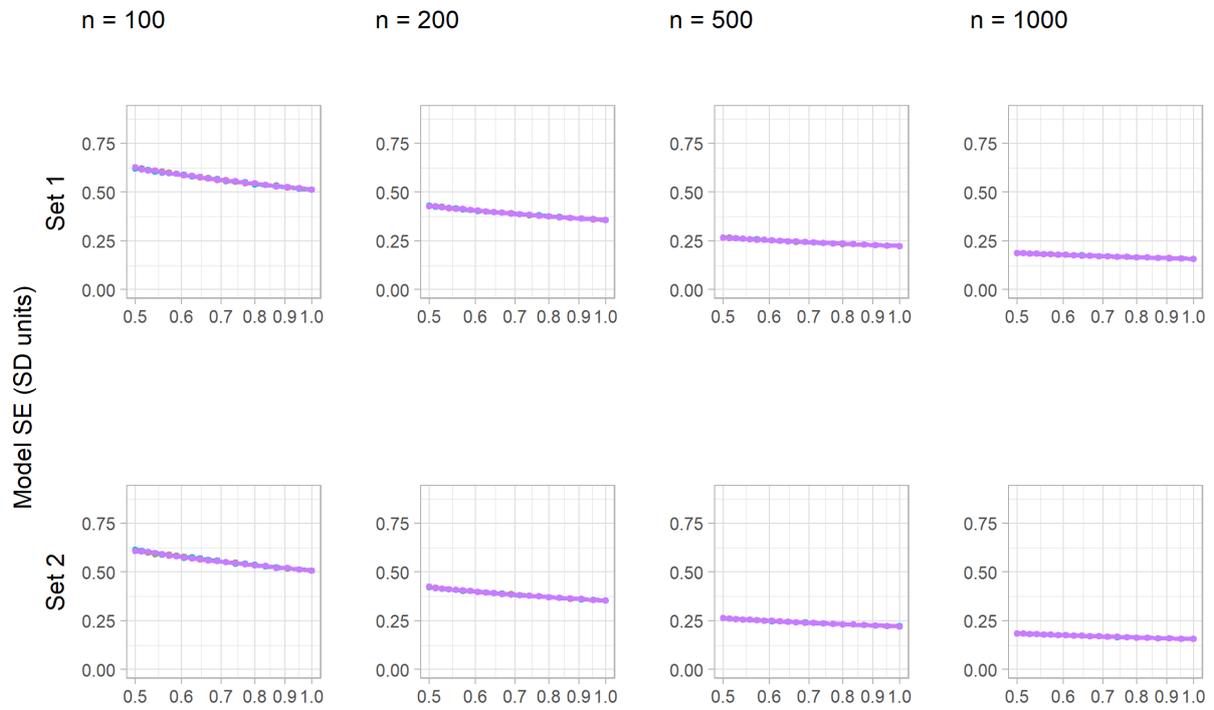

*S Figure 16: Model SE of a simple difference in means in the continuous outcome study (sensitivity analysis B), changed direction of effect on intermediate). Colour indicates treatment effect on the outcome variable (SDs): red = 0, green = 0.2, blue = 1, purple = 5.*

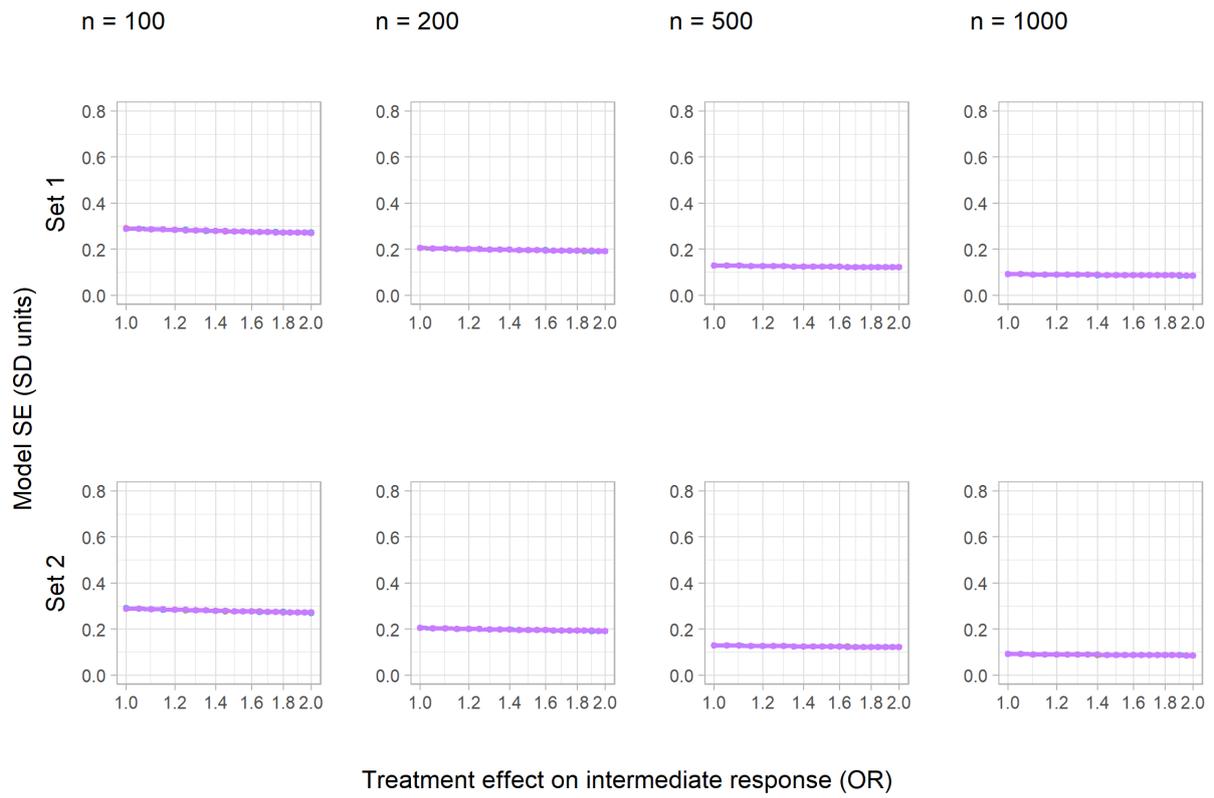

S Figure 17: Model SE of a simple difference in means in the continuous outcome study (sensitivity analysis C) increased event rate). Colour indicates treatment effect on the outcome variable (SDs): red = 0, green = 0.2, blue = 1, purple = 5.

# Binary outcomes
## Missing data

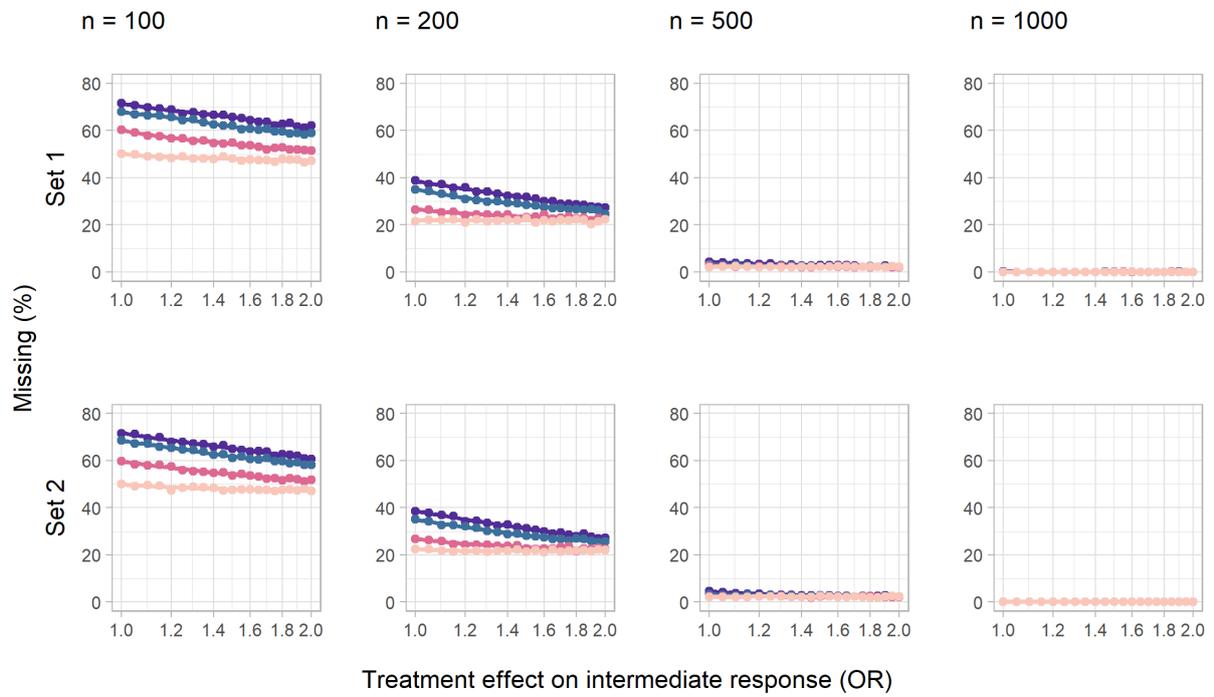

S Figure 18: Amount of missing data due to the treatment effect OR being inestimable in the binary outcome study (core scenarios). Colour indicates treatment effect on the outcome (ORs) (Purple = 1, blue = 1.2, darkpink = 2, light pink = 5)

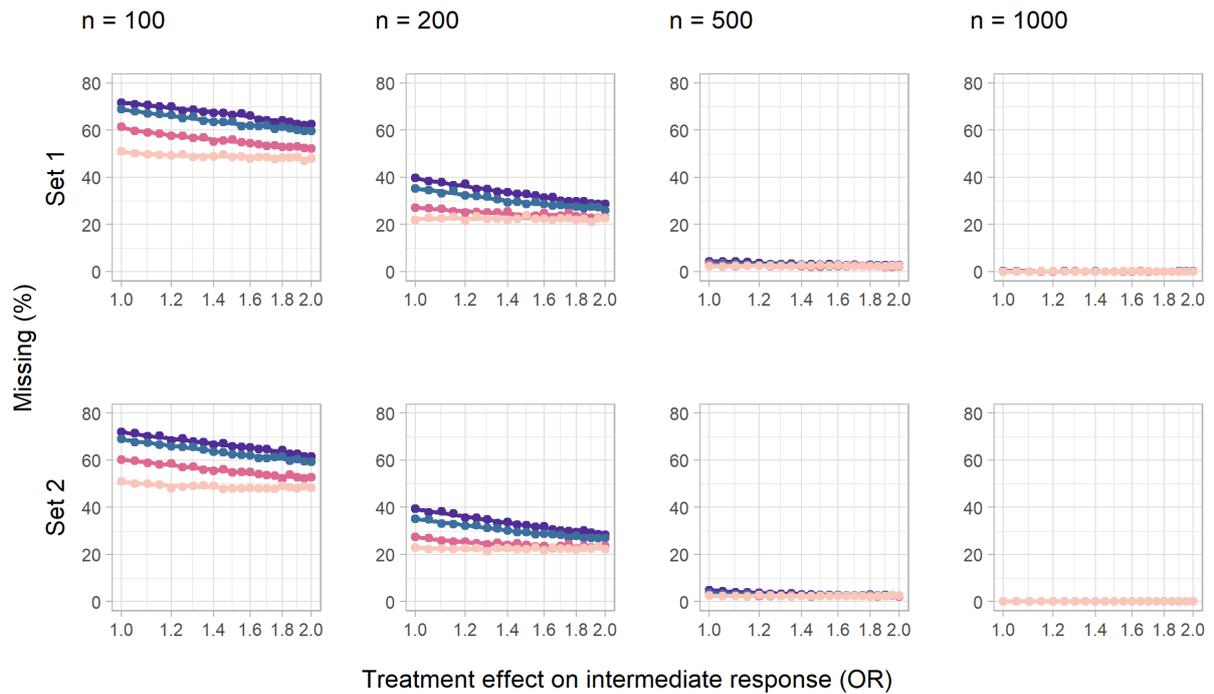

*S Figure 19: Amount of missing data due to the treatment effect OR being inestimable in the binary outcome study (sensitivity analysis A) increased confounding). Colour indicates treatment effect on the outcome (ORs) (Purple = 1, blue = 1.2, darkpink = 2, light pink = 5)*

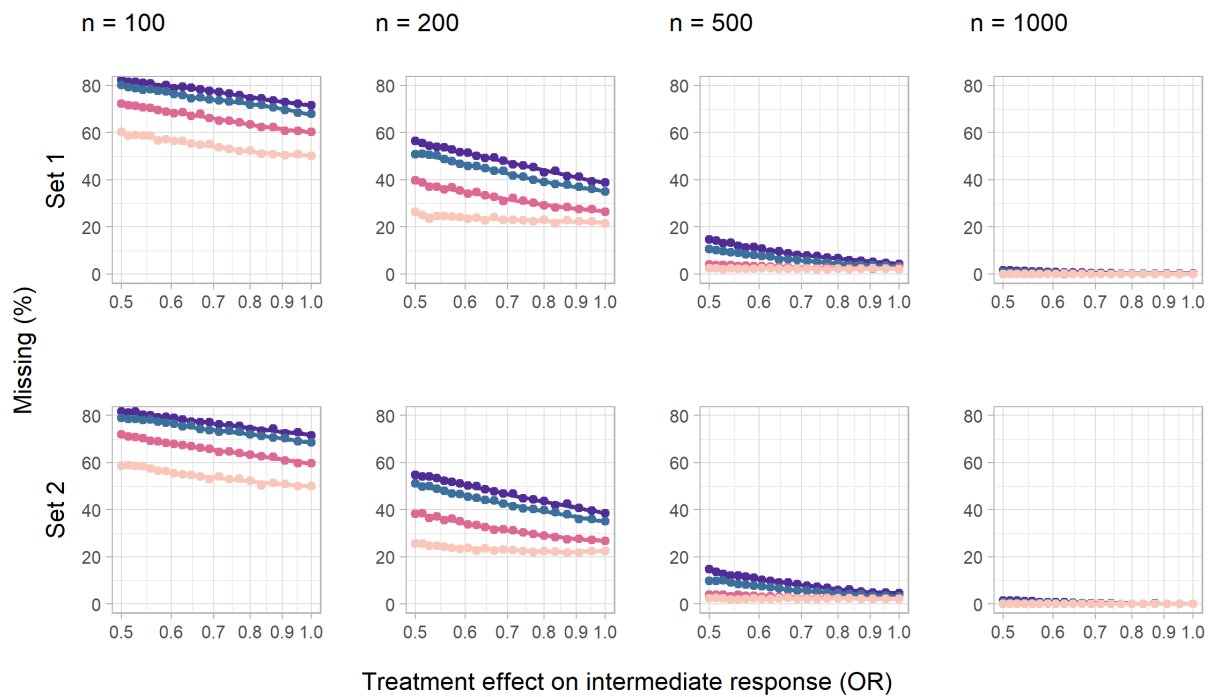

*S Figure 20: Amount of missing data due to the treatment effect OR being inestimable in the binary outcome study (sensitivity analysis B) changed sign of treatment effect on intermediate). Colour indicates treatment effect on the outcome (ORs) (Purple = 1, blue = 1.2, darkpink = 2, light pink = 5)*

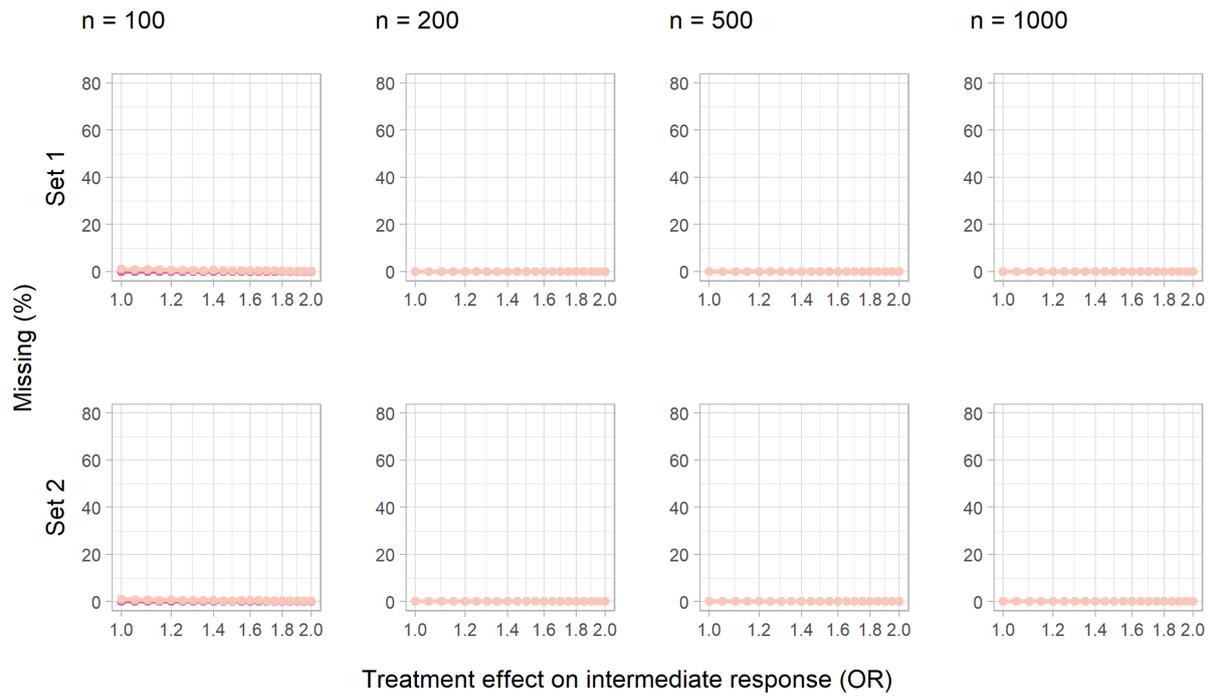

*S Figure 21: Amount of missing data due to the treatment effect OR being inestimable in the binary outcome study (sensitivity analysis C) increased event rate). Colour indicates treatment effect on the outcome (ORs) (Purple = 1, blue = 1.2, darkpink = 2, light pink = 5).*

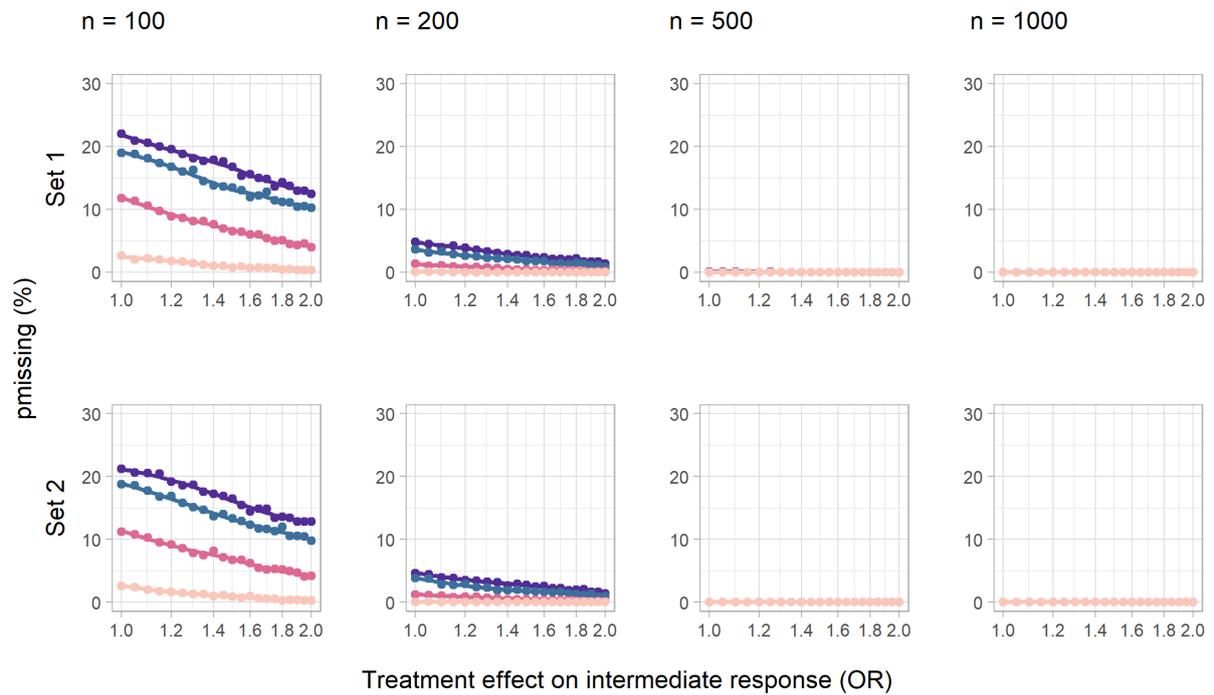

*S Figure 22: Amount of missing data due to the chi-squared statistic being incalculable in the binary outcome study (core scenarios). Colour indicates treatment effect on the outcome (ORs) (Purple = 1, blue = 1.2, darkpink = 2, light pink = 5)*

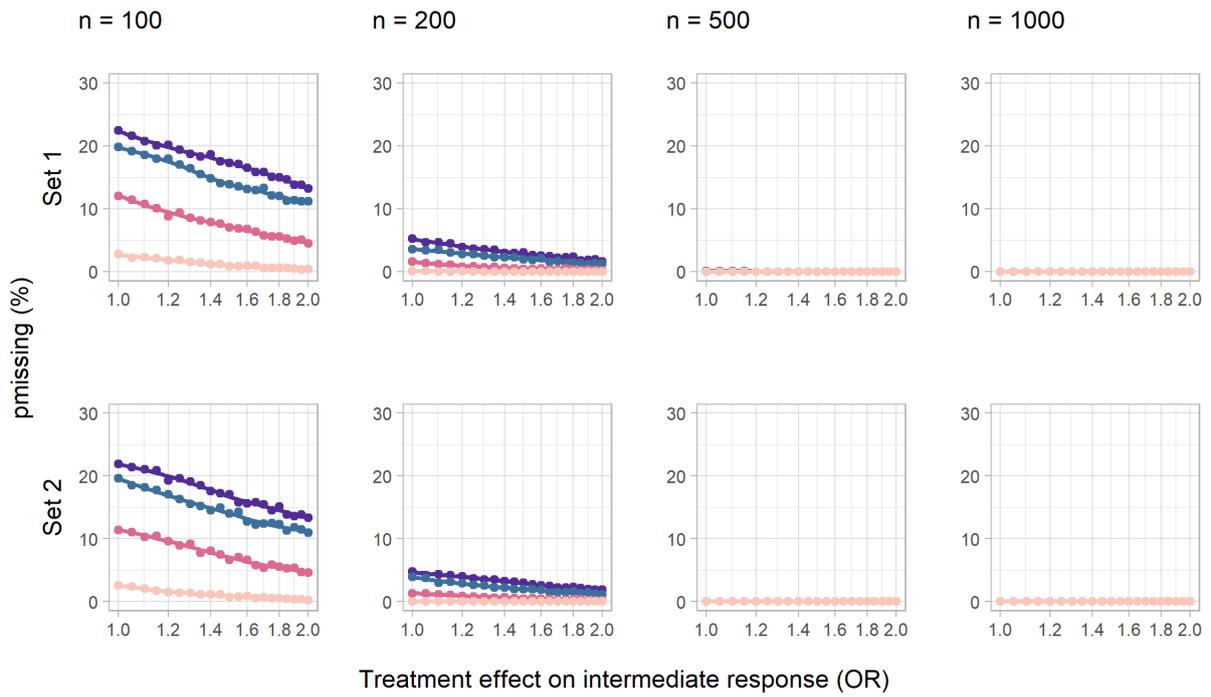

S Figure 23: Amount of missing data due to the chi-squared statistic being incalculable in the binary outcome study (sensitivity analysis A) increased confounding). Colour indicates treatment effect on the outcome (ORs) (Purple = 1, blue = 1.2, darkpink = 2, light pink = 5).

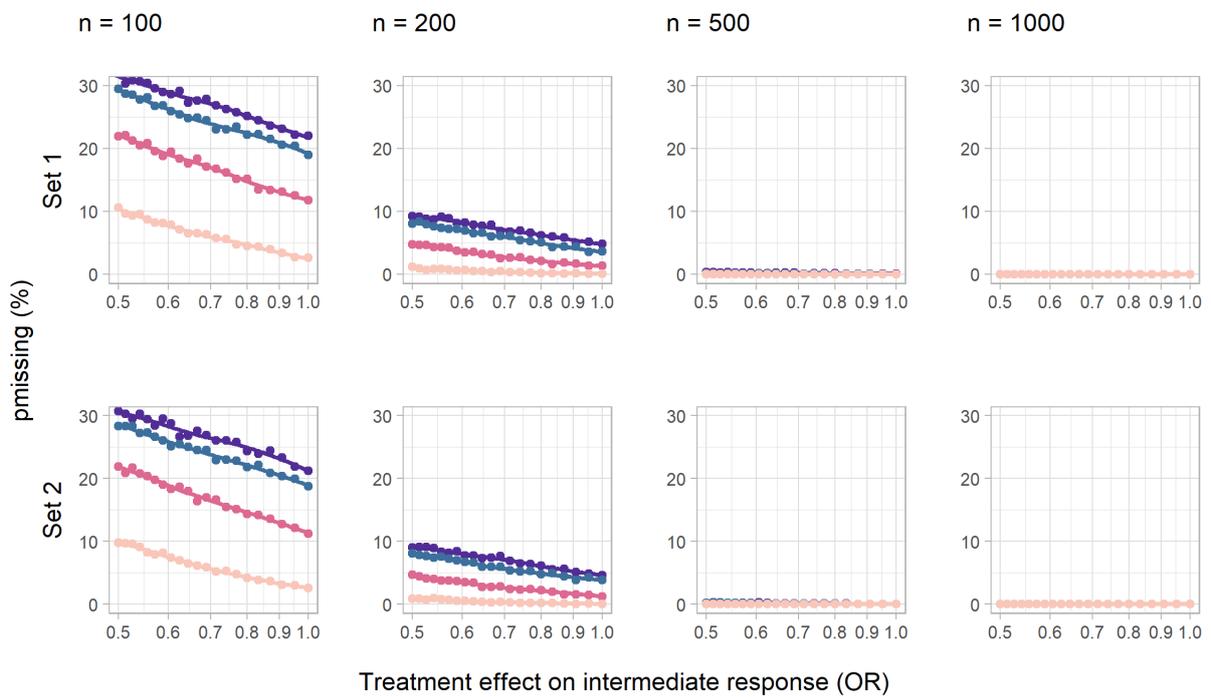

S Figure 24: Amount of missing data due to the chi-squared statistic being incalculable in the binary outcome study (sensitivity analysis B) changed direction of treatment effect on intermediate). Colour indicates treatment effect on the outcome (ORs) (Purple = 1, blue = 1.2, darkpink = 2, light pink = 5)

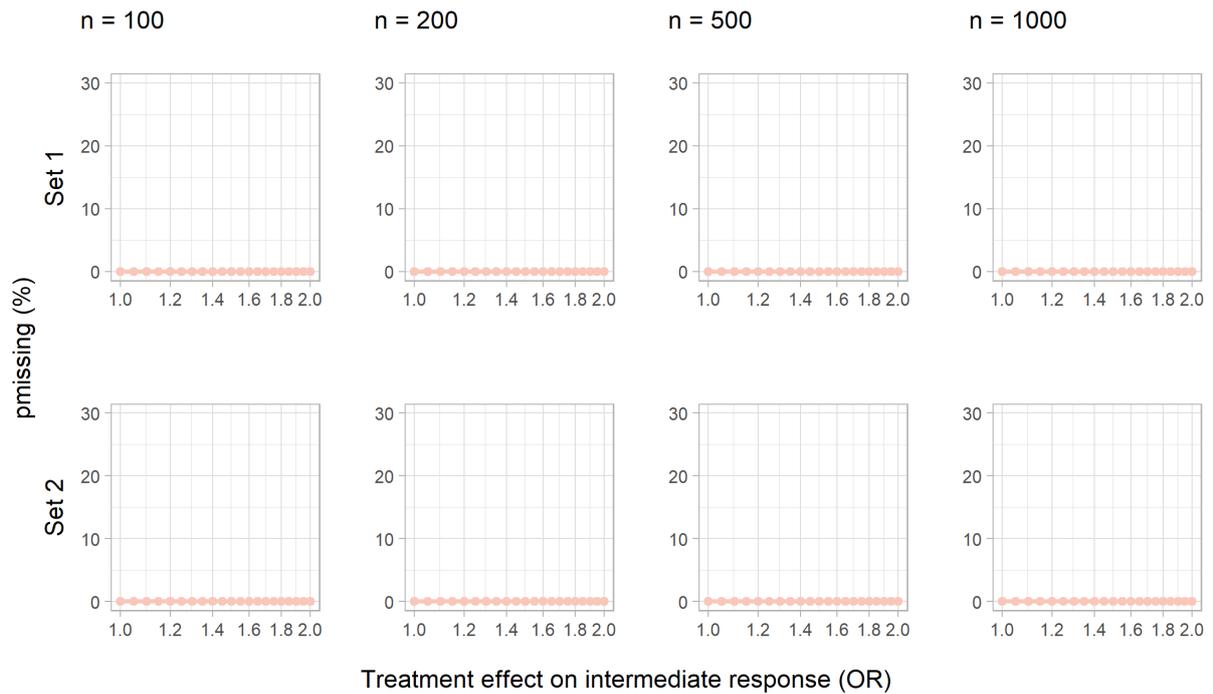

*S Figure 25: Amount of missing data due to the chi-squared statistic being incalculable in the binary outcome study (sensitivity analysis C) increased event rate). Colour indicates treatment effect on the outcome (ORs) (Purple = 1, blue = 1.2, darkpink = 2, light pink = 5)*

## Bias

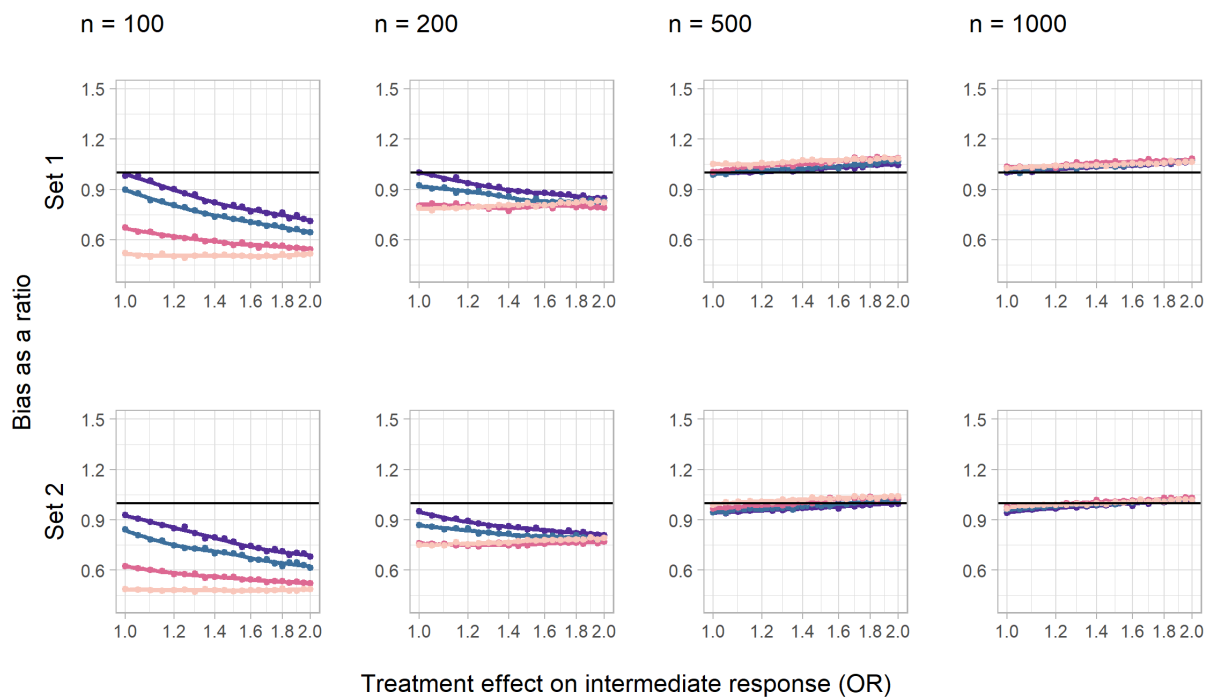

*S Figure 26: Bias expressed as a ratio of estimated to true odds ratio (ROR) in the binary outcome study (sensitivity analysis A), increased confounding). Colour indicates treatment effect on the outcome (ORs) (Purple = 1, blue = 1.2, darkpink = 2, light pink = 5)*

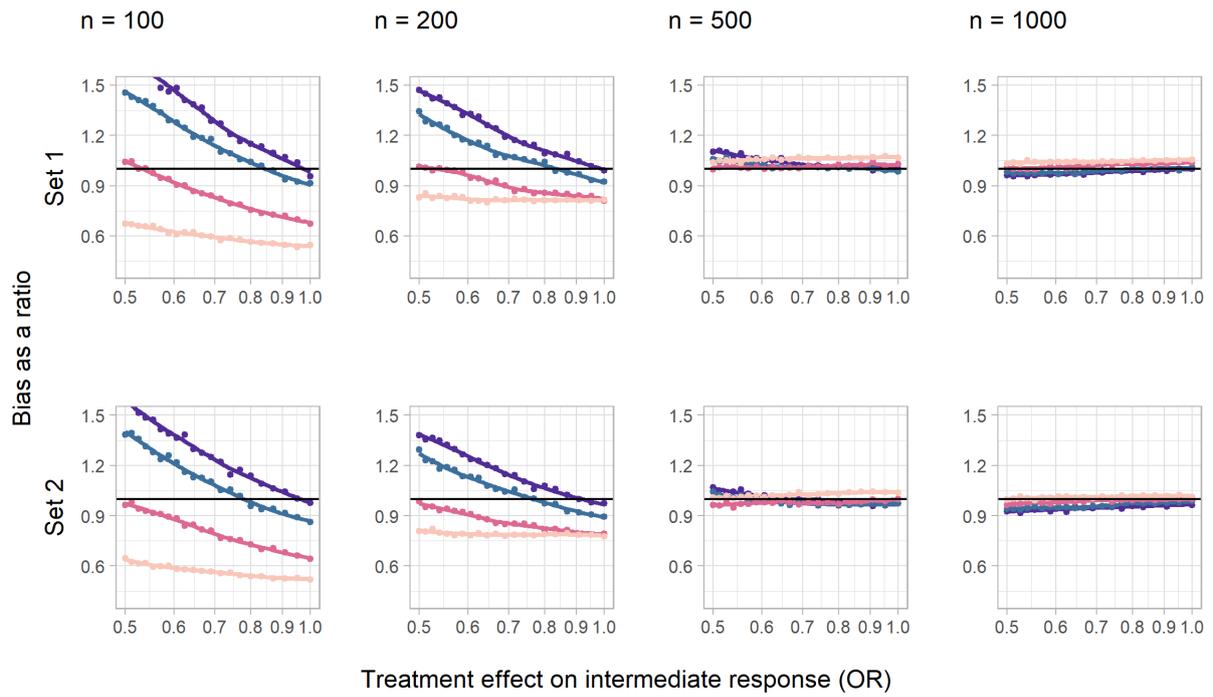

S Figure 27: Bias expressed as a ratio of estimated to true odds ratio (ROR) in the binary outcome study (sensitivity analysis B), changed direction of treatment effect on intermediate). Colour indicates treatment effect on the outcome (ORs) (Purple = 1, blue = 1.2, darkpink = 2, light pink = 5)

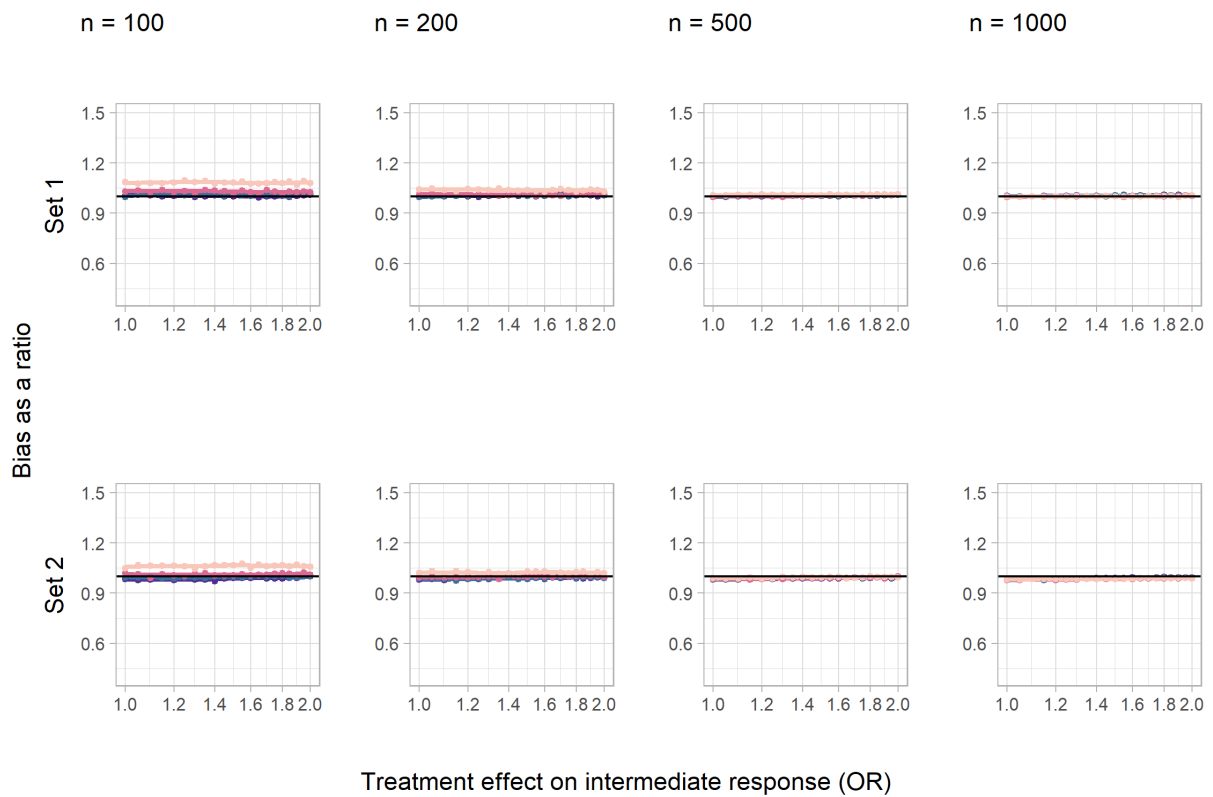

S Figure 28: Bias expressed as a ratio of estimated to true odds ratio (ROR) in the binary outcome study (sensitivity analysis C), increased event rate). Colour indicates treatment effect on the outcome (ORs) (Purple = 1, blue = 1.2, darkpink = 2, light pink = 5)

Coverage

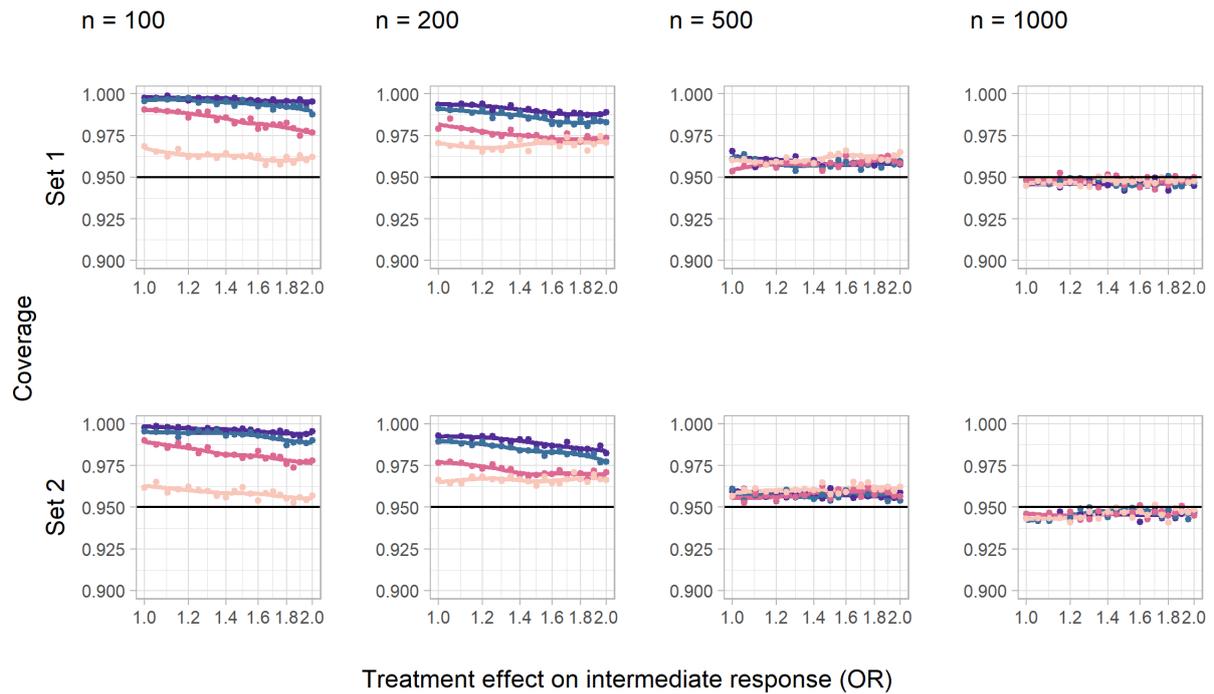

S Figure 29: Coverage of 95% confidence interval obtained using logistic regression in the binary outcome study (sensitivity analysis A) increased confounding). Colour indicates treatment effect on the outcome (ORs) (Purple = 1, blue = 1.2, darkpink = 2, light pink = 5).

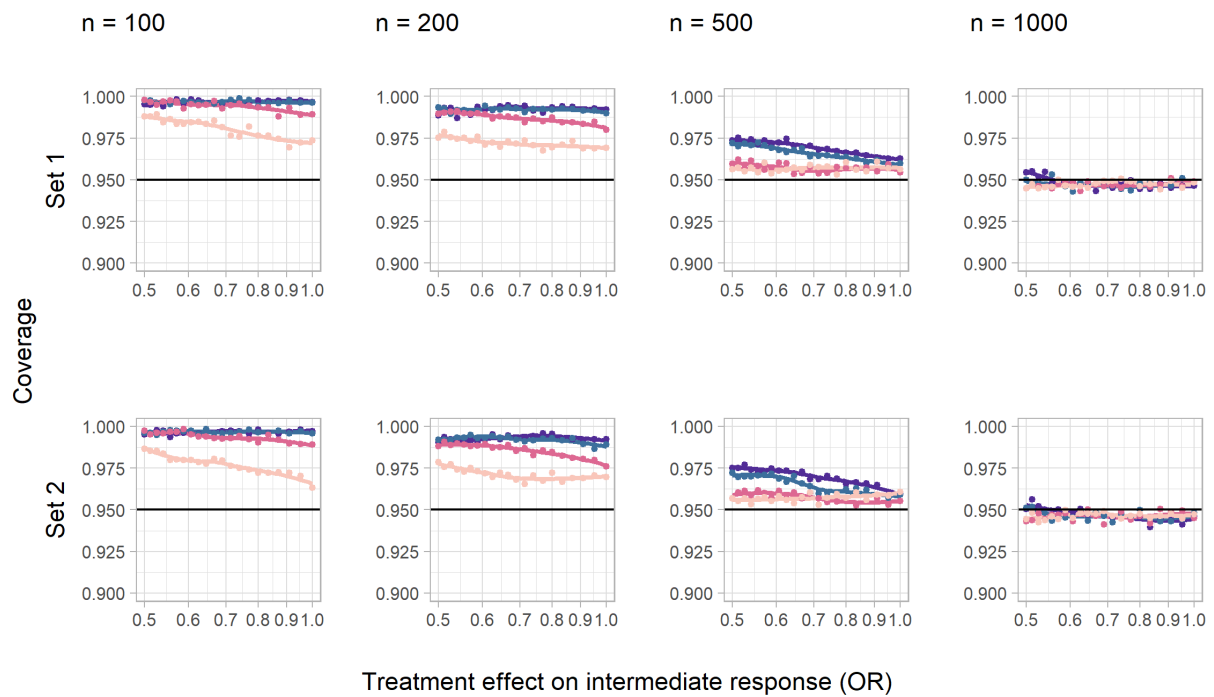

S Figure 30: Coverage of 95% confidence interval obtained using logistic regression in the binary outcome study (sensitivity analysis B) changed direction of treatment effect on intermediate). Colour indicates treatment effect on the outcome (ORs) (Purple = 1, blue = 1.2, darkpink = 2, light pink = 5).

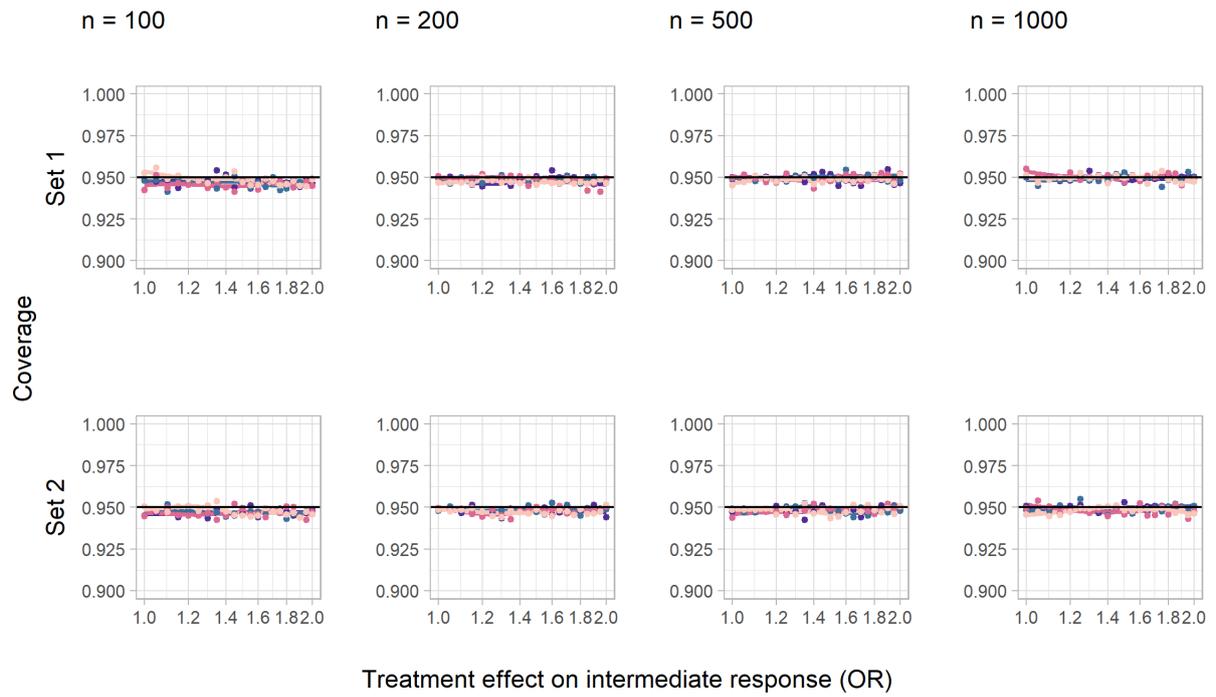

*S Figure 31: Coverage of 95% confidence interval obtained using logistic regression in the binary outcome study (sensitivity analysis C) increased event rate). Colour indicates treatment effect on the outcome (ORs) (Purple = 1, blue = 1.2, darkpink = 2, light pink = 5).*

# Empirical SE

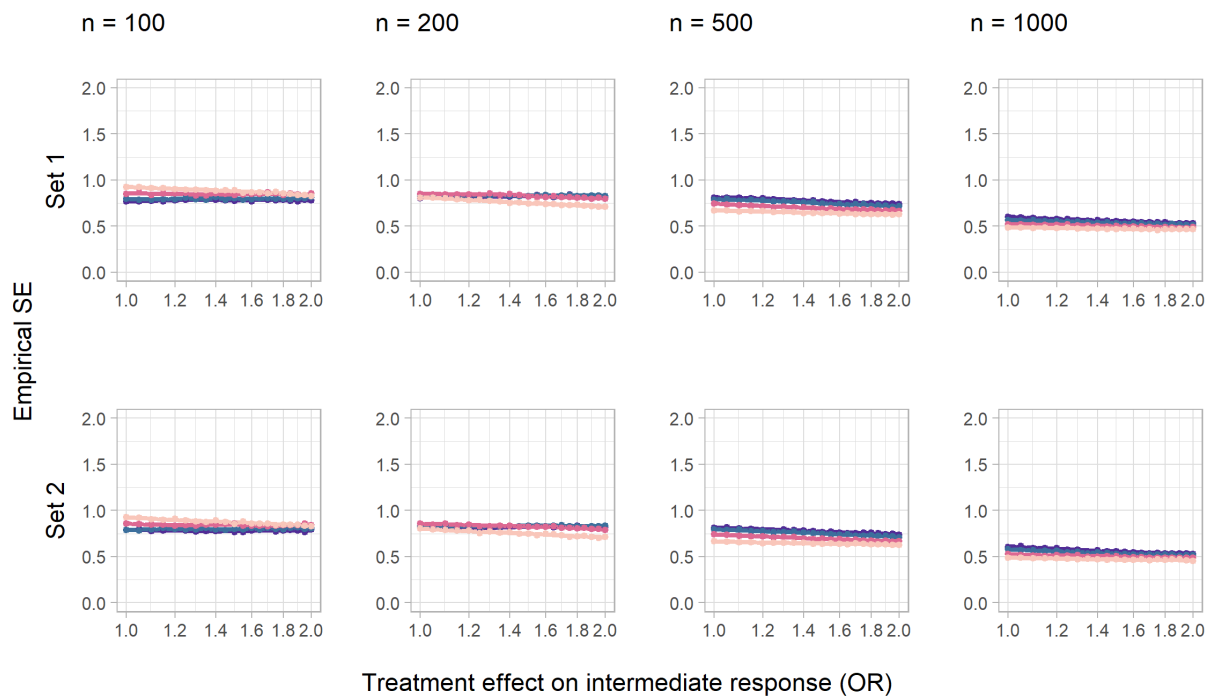

S Figure 32: Empirical SE in the binary outcome study (core scenarios). Colour indicates treatment effect on the outcome (ORs) (Purple = 1, blue = 1.2, darkpink = 2, light pink = 5).

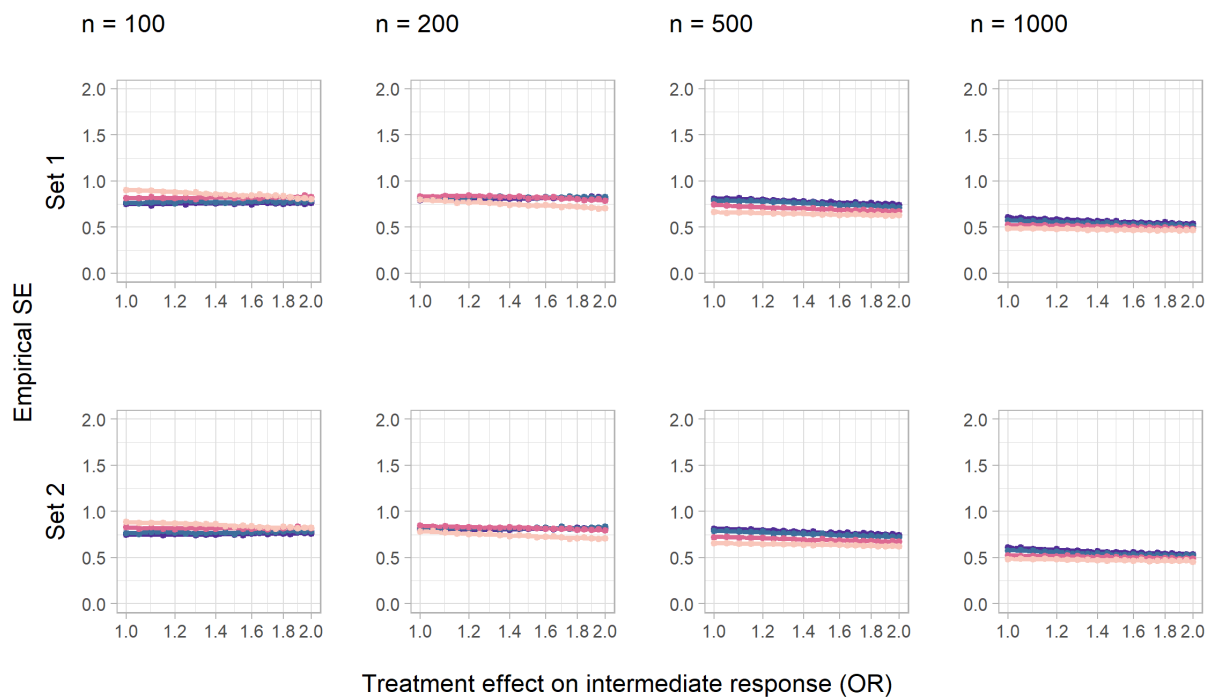

S Figure 33: Empirical SE in the binary outcome study (sensitivity analysis A) increased confounding). Colour indicates treatment effect on the outcome (ORs) (Purple = 1, blue = 1.2, darkpink = 2, light pink = 5).

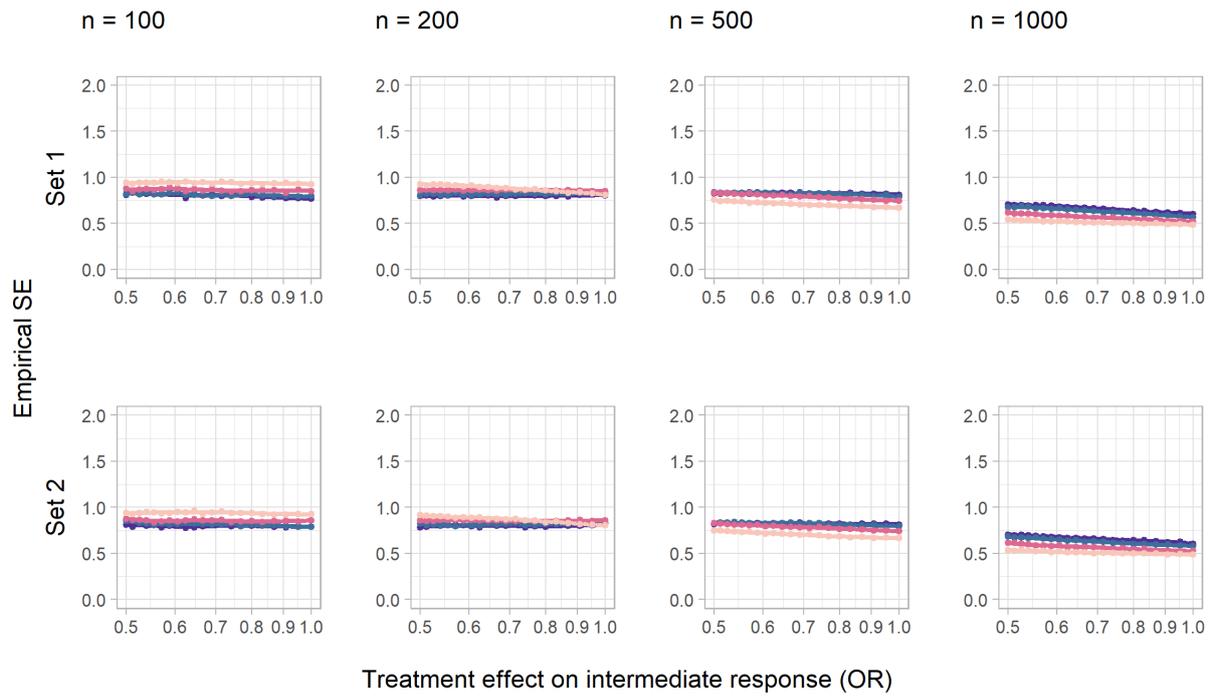

*S Figure 34: Empirical SE in the binary outcome study (sensitivity analysis B) changed sign of treatment effect on intermediate variable). Colour indicates treatment effect on the outcome (ORs) (Purple = 1, blue = 1.2, darkpink = 2, light pink = 5).*

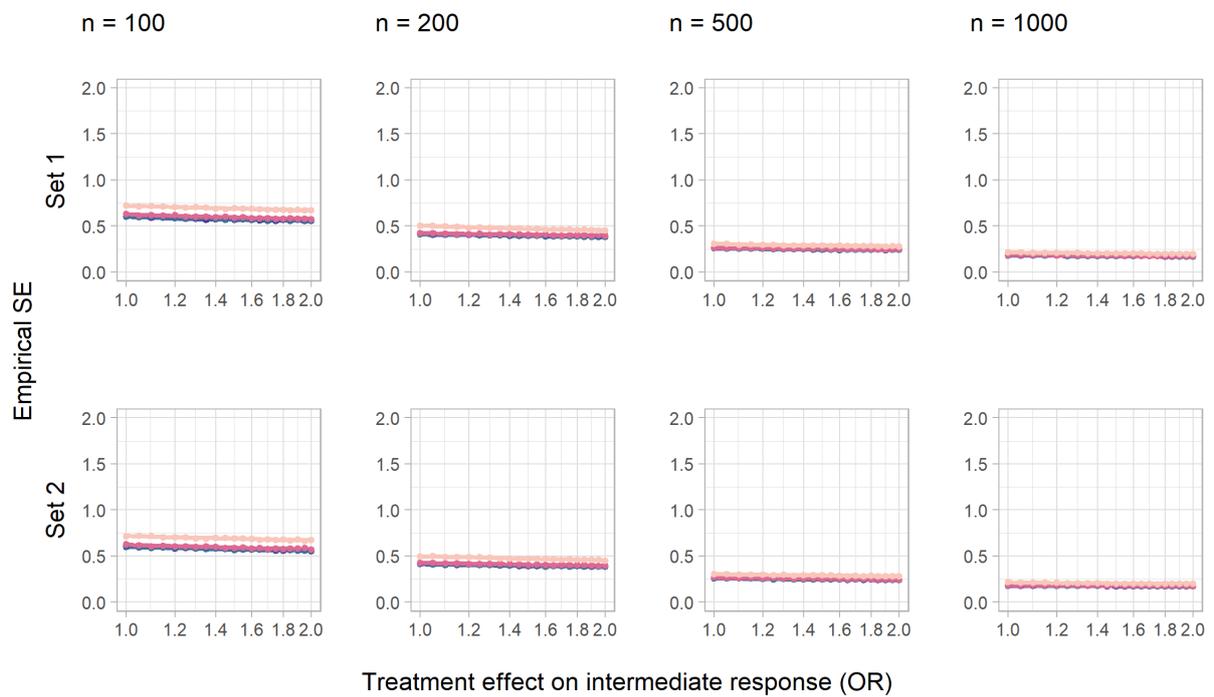

*S Figure 35: Empirical SE in the binary outcome study (sensitivity analysis C) increased event rate). Colour indicates treatment effect on the outcome (ORs) (Purple = 1, blue = 1.2, darkpink = 2, light pink = 5).*

# Model SE

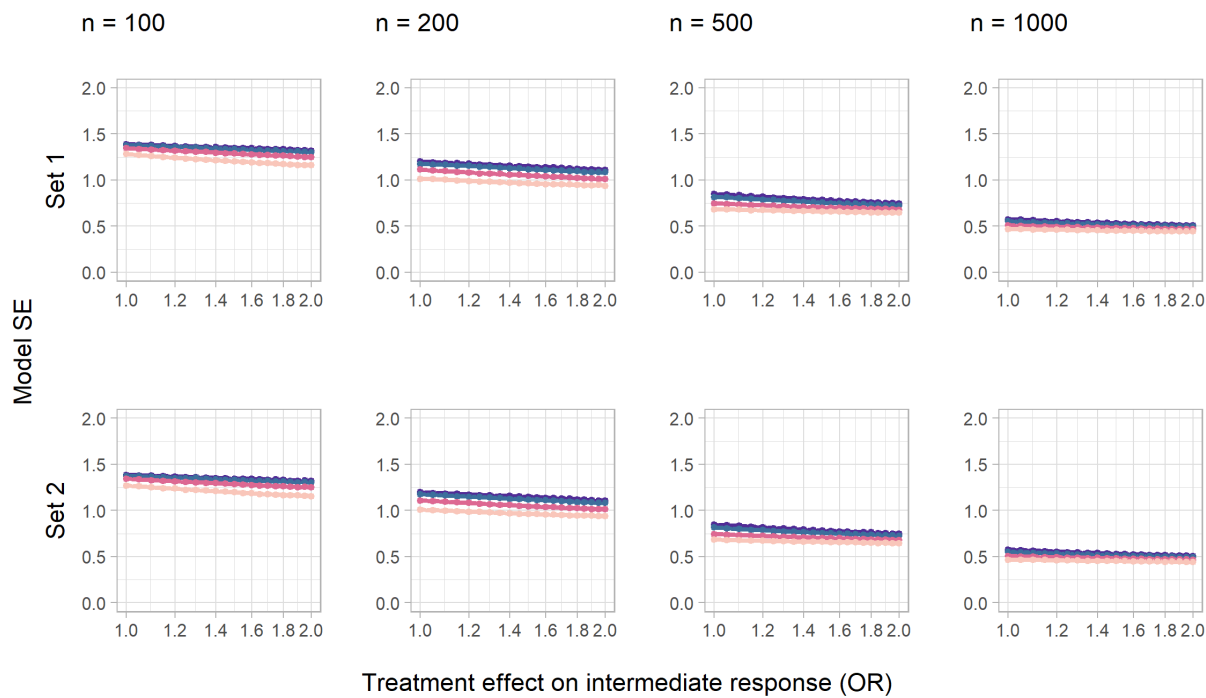

*S Figure 36: Model SE in the binary outcome study (core scenarios). Colour indicates treatment effect on the outcome (ORs) (Purple = 1, blue = 1.2, darkpink = 2, light pink = 5).*

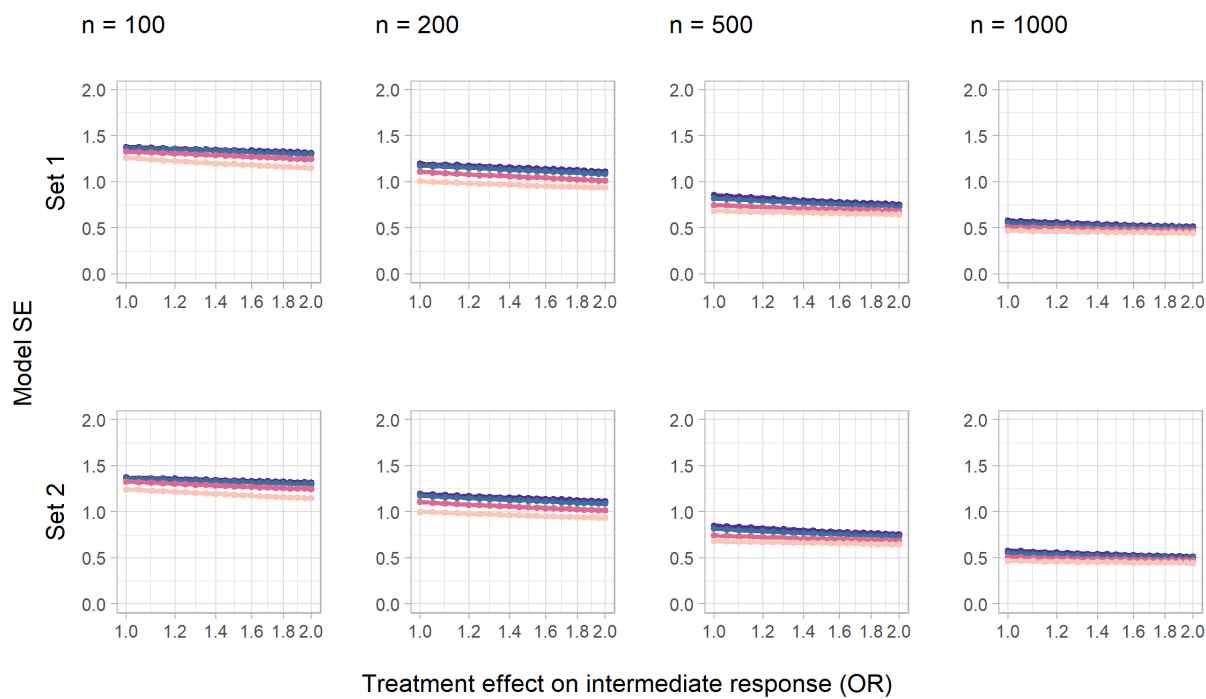

*S Figure 37: Model SE in the binary outcome study (sensitivity analysis A) increased confounding). Colour indicates treatment effect on the outcome (ORs) (Purple = 1, blue = 1.2, darkpink = 2, light pink = 5).*

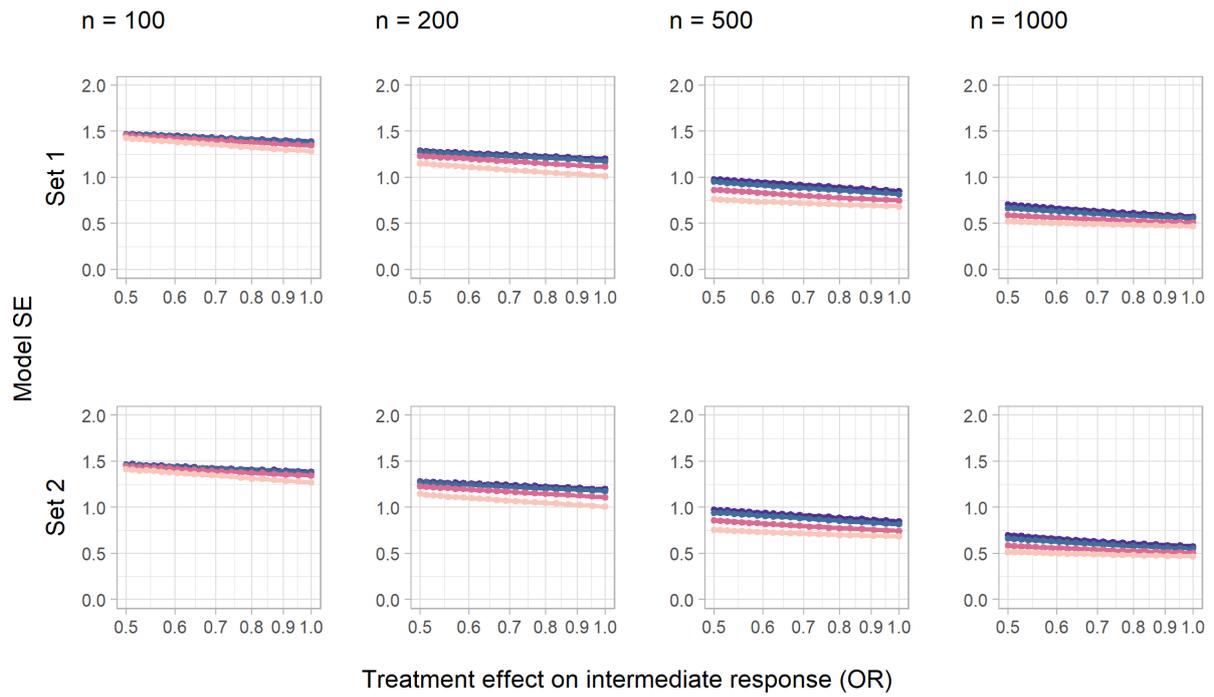

*S Figure 38: Model SE in the binary outcome study (sensitivity analysis B) changed sign of effect on intermediate variable). Colour indicates treatment effect on the outcome (ORs) (Purple = 1, blue = 1.2, darkpink = 2, light pink = 5).*

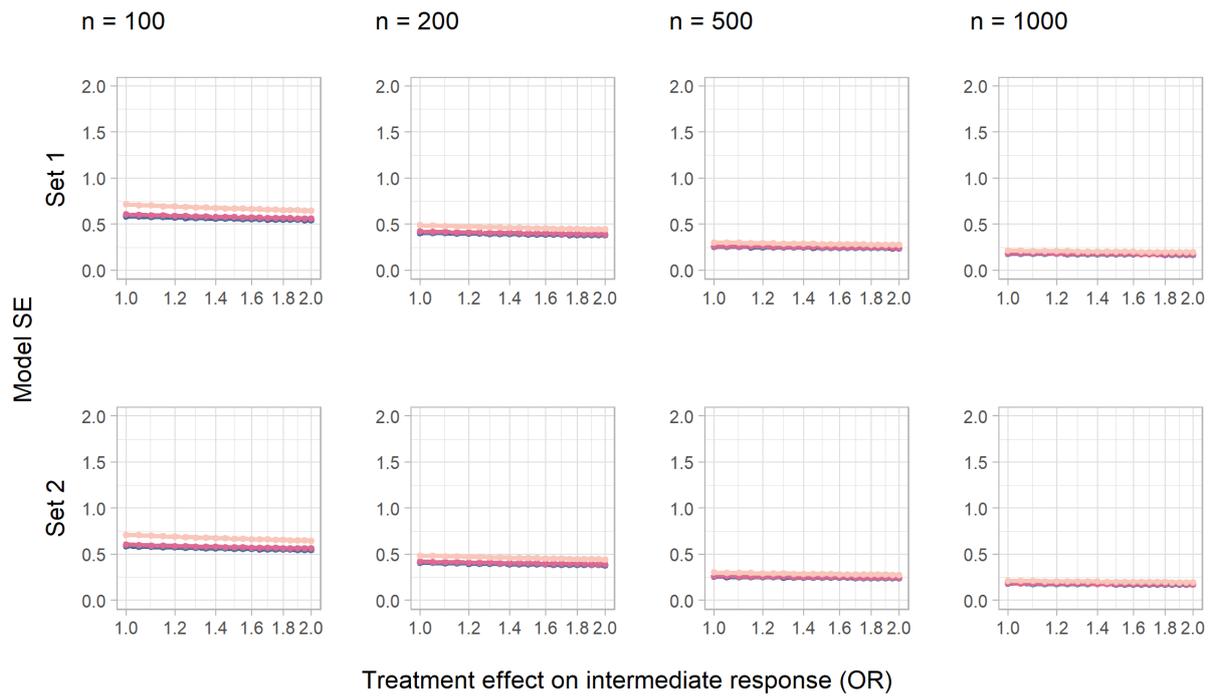

*S Figure 39: Model SE in the binary outcome study (sensitivity analysis C) increased event rate). Colour indicates treatment effect on the outcome (ORs) (Purple = 1, blue = 1.2, darkpink = 2, light pink = 5).*

## Type 1 error

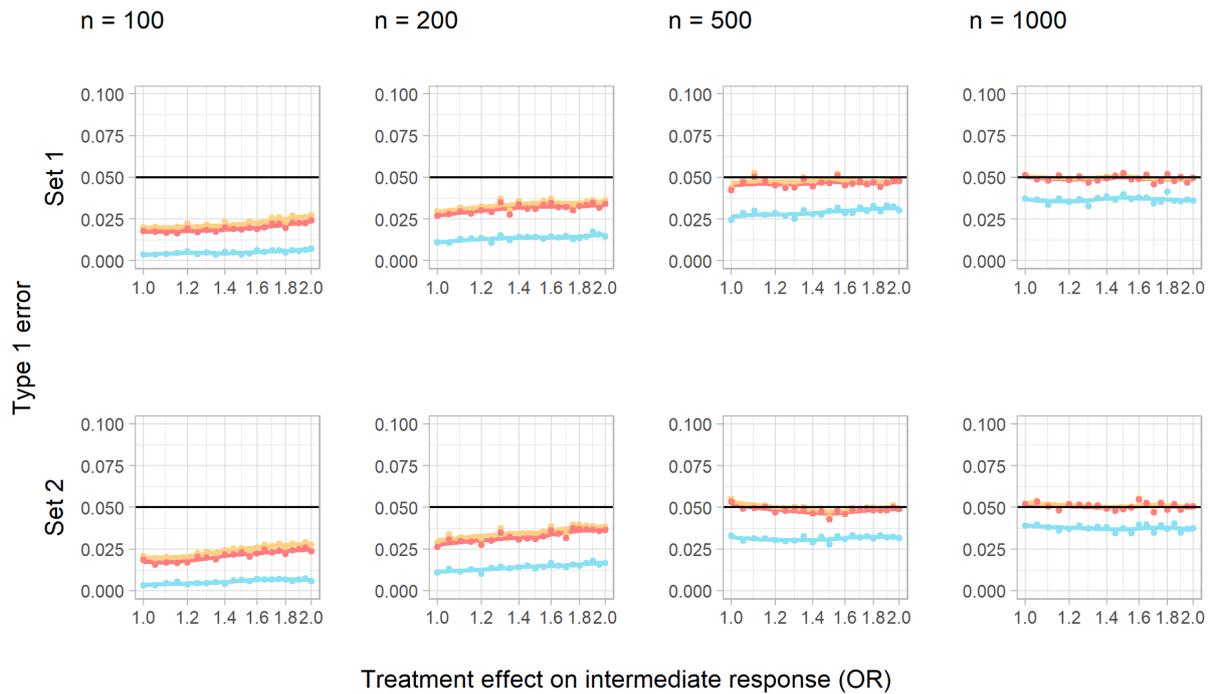

*S Figure 40: Type 1 error of statistical tests in the binary outcome study (sensitivity analysis A) increased confounding). Fisher's exact test = blue, chi-squared test = pink, adjusted chi-squared test = yellow.*

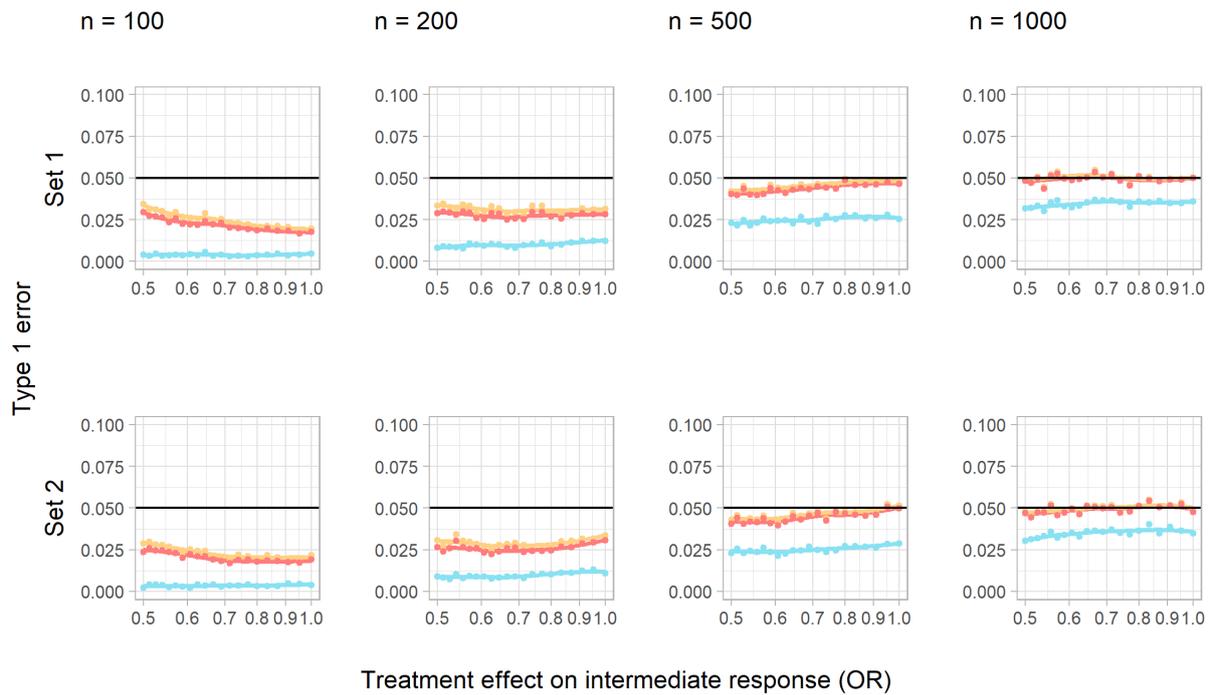

*S Figure 41: Type 1 error of statistical tests in the binary outcome study (sensitivity analysis B) changed direction of treatment effect on intermediate). Fisher's exact test = blue, chi-squared test = pink, adjusted chi-squared test = yellow.*

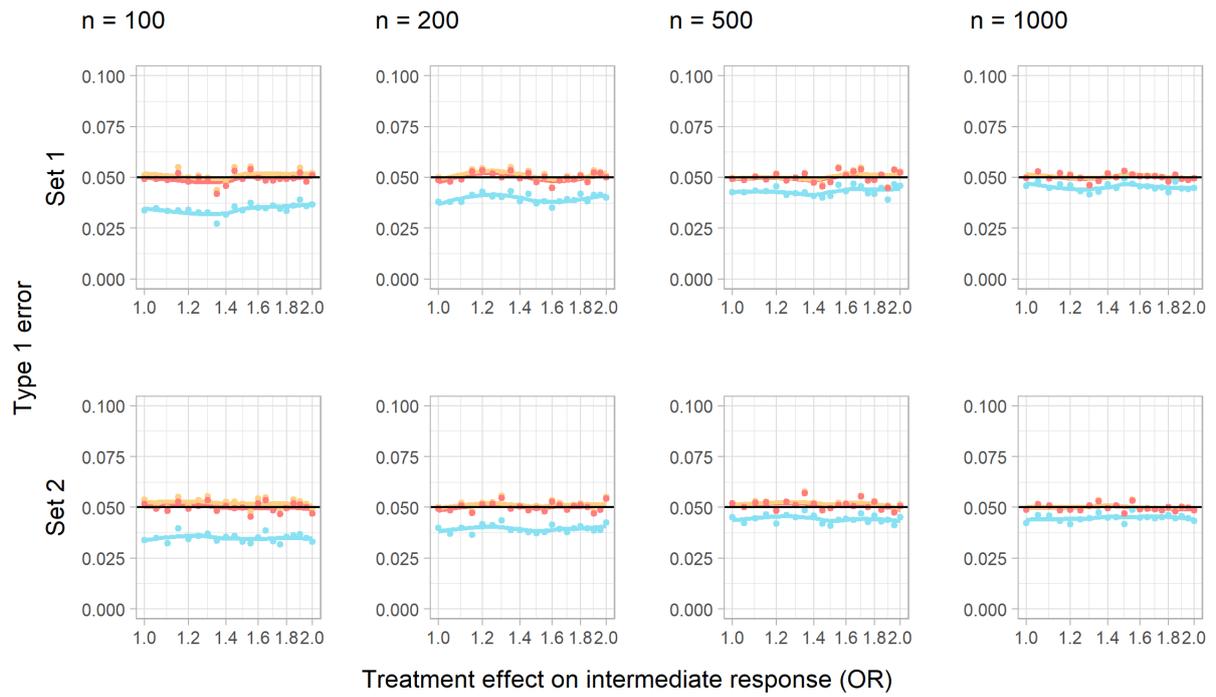

*S Figure 42: Type 1 error of statistical tests in the binary outcome study (sensitivity analysis C) increased event rate). Fisher's exact test = blue, chi-squared test = pink, adjusted chi-squared test = yellow*